\documentclass[a4paper,11pt]{article}
\pdfoutput=1

\usepackage{jheppub}
\usepackage{natbib}
\usepackage[utf8]{inputenc}
\usepackage{mathabx}
\usepackage{subcaption}
\usepackage[title]{appendix}

\newcommand{\seff}{S_{\textrm{eff}}}
\newcommand{\kgc}{k_0^c}
\newcommand{\kcc}{k_3^c}
\newcommand{\optwo}{O_2}
\newcommand{\tn}{\tilde{N}_{31}}
\newcommand{\avn}{\langle n_t \rangle}
\newcommand{\aavn}{\langle n \rangle}
\newcommand{\bn}{\bar{n}}

\graphicspath{ {./images/} }

\title{The phase structure and effective action of 3D CDT at higher spatial genus}
\author[a]{Joren Brunekreef}
\author[b]{Dániel Németh}

\affiliation[a]{Institute for Mathematics, Astrophysics and Particle Physics, Radboud University, \\
Heyendaalseweg 135, 6525 AJ Nijmegen, The Netherlands.}
\affiliation[b]{Institute of Theoretical Physics, Jagiellonian University, \\
Łojasiewicza 11, Kraków, PL 30-348, Poland.}

\emailAdd{jorenb@gmail.com}
\emailAdd{nemeth.daniel.1992@gmail.com}

\date{\today}

\abstract{We perform a detailed investigation of the phase structure and the semiclassical effective action of (2+1)-dimensional Causal Dynamical Triangulations (CDT) quantum gravity using computer simulations. On the one hand, we study the effect of enlarging the ensemble of triangulations by relaxing the simplicial manifold conditions in a controlled way. On the other hand, we cast a first look at CDT geometries with spatial topology beyond that of the sphere or torus. We measure the phase structure of the model for several triangulation ensembles and spatial topologies, finding evidence that the phase structure is qualitatively unaffected by these generalizations. Furthermore, we determine the effective action for the spatial volumes of the system, again varying the simplicial manifold conditions and the spatial topology. In all cases where we were able to gather sufficient statistics, we found the resulting effective action to be consistent with a minisuperspace action derived from continuum Einstein gravity. We interpret our overall results as evidence that 1) partially relaxing simplicial manifold conditions or changing the spatial genus does not affect the continuum limit of 3D CDT and that 2) increasing the spatial genus of the system likely does not influence the leading-order terms in the emergent effective action.}

\begin{document}
\maketitle

\newpage
\section{Introduction}
A candidate approach for tackling the notoriously difficult problem of quantizing gravity is the framework of Causal Dynamical Triangulations (CDT) \cite{ambjorn2012nonperturbative,loll2020quantum}. The starting point for this model is the formal quantum gravitational path integral, which for pure gravity in $D$ dimensions reads
\begin{equation}
    \mathcal{Z} = \int \mathcal{D}[g] \, e^{i S^\textrm{EH}[g]},
    \quad \quad
    S^\textrm{EH}[g] = \frac{1}{16 \pi G} \int d^D x \sqrt{-\det g} (R-2\Lambda).
    \label{eq:qg-pi}
\end{equation}
The integration is over the space of diffeomorphism-equivalence classes of Lorentzian metrics $g_{\mu \nu}(x)$ on a fixed $D$-dimensional manifold $M$, and $S^\textrm{EH}$ is the Einstein-Hilbert action with Einstein gravitational constant $G$ and cosmological constant $\Lambda$. This purely formal expression is defined in the CDT approach as the continuum limit of a regularized counterpart of the path integral, where the space of Lorentzian metrics (modulo diffeomorphisms) is approximated as a space of piecewise flat Lorentzian manifolds with a causal structure. The setup advocated in CDT is similar in spirit to non-perturbative treatments of quantum chromodynamics (QCD) on a discrete spacetime lattice.

One of the important features of CDT is the presence of a well-defined Wick rotation that maps the regularized Lorentzian path integral to a Euclidean one, which can be understood as the partition function of a statistical system of \emph{triangulated geometries}. This makes the model solvable analytically for $D=2$ \cite{ambjorn1998nonperturbative}, and allows for a numerical treatment in $D=2,3,4$ \cite{ambjorn2001nonperturbative,ambjorn2001dynamically} using Monte Carlo methods. Furthermore, the regularized path integral provides a well-defined notion of a space of geometries, which sidesteps having to deal with difficulties related to gauge-fixing the diffeomorphism symmetry of gravity.

The version of the CDT model with $D=4$ is a promising route towards understanding fully-fledged four-dimensional quantum gravity. Monte Carlo simulations of this system have indicated the presence of a second-order phase transition \cite{ambjorn2011secondorder,ambjorn2012second}, which is an essential ingredient for constructing a continuum limit of the regularized theory in the presence of local propagating degrees of freedom. Furthermore, the effective geometry that emerges from this model resembles a four-dimensional de Sitter universe in several respects on sufficiently coarse-grained scales \cite{ambjorn2004emergence,ambjorn2021cdt,ambjorn2008planckian,ambjorn2008nonperturbative,klitgaard2020how}. The historical origins of the CDT approach can be found in the study of the \emph{Euclidean} version of the quantum gravitational path integral from a regularized point of view, a framework which is referred to as Euclidean Dynamical Triangulations (EDT or simply DT). This approach, in turn, can be traced back to early attempts to quantize the bosonic string \cite{david1985planar,kazakov1985critical,durhuus1984critical,ambjorn1997quantumb}. Subsequent investigations of the triangulated Euclidean path integral in higher dimensions have failed to demonstrate the existence of a higher-order phase transition, making it difficult to see how a continuum theory of quantum gravity can emerge from these models. This suggests that integrating over the space of Lorentzian vs. Euclidean spacetimes leads to genuinely different results, and perhaps that enforcing a notion of causality (which is not present in DT) is essential for making the path integral well-behaved.

The CDT simplicial manifolds can be foliated into a series of spatial hypersurfaces $\Sigma_t$ of equal topology, where $t$ is a discrete label corresponding to a notion of proper time. The full simplicial manifold $\mathcal{M}$ has the product topology $\Sigma_t \times [0,1]$. A useful observable characterizing the overall shape of a CDT geometry is its \emph{volume profile}, which measures the size of each spatial `universe' $\Sigma_t$ as a function of the time label $t$. It was found \cite{ambjorn2004emergence} that the average volume profile in a certain phase of 4D CDT of $S^3 \times S^1$ topology\footnote{The `time' direction is typically given periodic boundary conditions, resulting in the $S^1$ topology.} (where $S^n$ is an $n$-sphere) can be matched with good accuracy to that of a classical 4D de Sitter universe. What is more, it turns out that the quantum fluctuations around this average `background' universe are very well described by a semiclassical \emph{effective action} for the spatial volume \cite{ambjorn2005semiclassical,ambjorn2005reconstructing,ambjorn2008planckian}, and that this effective action is consistent with a discretized version of the minisuperspace action derived from the Einstein-Hilbert action for continuum 4D gravity.

The focus of this work will be CDT quantum gravity in three dimensions \cite{ambjorn2001nonperturbative}. This nonperturbative model of (2+1)-dimensional Lorentzian quantum gravity \cite{carlip1998quantum} is a toy model of bona fide (3+1)-dimensional quantum gravity. Such a toy model can serve to deepen our understanding of the techniques used in the simplicial context, and possibly to help to interpret observations made in the physically more interesting four-dimensional system. It was found in \cite{ambjorn1998nonperturbative} using numerical simulations that this three-dimensional toy model has an interesting two-phase structure: in one region of its one-dimensional parameter space (the \emph{degenerate} phase), the volume profiles fluctuate wildly without any trace of an effective background geometry, whereas in the other region (the \emph{de Sitter} phase) the volumes of neighboring spatial universes tend to align, and an extended three-dimensional object emerges. However, unlike the celebrated higher-order transition of CDT in four dimensions, the transition in 3D CDT is of first order. However, since gravity in three dimensions has no propagating degrees of freedom, such a higher-order transition is not required to formulate a continuum limit.

Furthermore, the volume profiles of this model with $S^2 \times S^1$ topology can be matched to a 3D de Sitter universe, analogous to the previously mentioned results in the 4D setting. This investigation was subsequently extended \cite{budd2012effective,budd2013exploring} to 3D CDT where the spatial universes are of \emph{torus} topology $\mathbb{T}^2$, so that the topology of the full spacetime is $\mathbb{T}^2 \times S^1$. Again, the results indicated that a discretized minisuperspace action is appropriate for describing both the average volume profile and the fluctuations around it. 

Our aim is to generalize these results to systems of 3D CDT where the spatial geometries are orientable closed surfaces with topologies different from $S^2$ or $\mathbb{T}^2$. The topology of such surfaces is fully characterized by their \emph{genus}, denoted $g$, a nonnegative integer. The spatial universes in our setup therefore correspond to $g$-holed tori, denoted $\mathbb{T}^g$. We intend to investigate whether it is still possible to construct effective actions for these systems with higher spatial genus, and to determine whether increasing the genus leads to identifiable correction terms in such an action.

As a precursor to determining the effective action, we make an in-depth study of the phase diagram of 3D CDT. On the one hand, we investigate whether the choice of spatial genus affects the phase structure and the nature of the transition. Furthermore, we focus on the possible effects of relaxing the \emph{simplicial manifold conditions} that are present in standard CDT. Usually, the CDT partition sum is taken over strict simplicial manifolds without allowing for local degeneracies in the triangulations. It was found in \cite{ambjorn2001lorentzian} that the degenerate phase disappears completely when several of the simplicial manifold conditions are dropped\footnote{It should be pointed out that the cited work makes use of \emph{quadrangulations} in the spatial hypersurfaces instead of triangulations, but it does not affect continuum behavior of the model.}, which the authors interpret as evidence that this degenerate phase is nothing more than a discretization artefact. In the present work, we make use of less drastic generalizations of the CDT ensemble, instead dropping simplicial manifold conditions in small steps and subsequently studying the effect hereof on the phase diagram. Using generalized ensembles may be of help in approaching continuum behavior at smaller system volumes compared to the strict simplicial manifold ensembles, as is known from \cite{ambjorn1995new} for the case of 2D DT. It has been shown that the continuum limit of the degenerate ensemble for this system of 2D DT leads to the same \emph{universal} behavior as when using strict 2D triangulations \cite{tutte1962census,brezin1978planar}, however no such universality result is known for 3D CDT. Studying the effect of relaxing simplicial manifold conditions in this model is therefore interesting from a practical point of view, since generalized ensembles may be used in place of strict ones if it can be made plausible that their behavior in the continuum limit is identical.

The structure of this work is the following: In the next section, we review the main ingredients of 3D CDT quantum gravity.\footnote{This work was carried out in parallel with \cite{brunekreef2022nature}, and we partly follow the exposition presented therein.} In Sec. \ref{sec:ensembles}, we define four distinct ensembles of triangulations that can be used in the CDT partition sum, where each ensemble is characterized by a set of conditions on the dual graph of the two-dimensional spatial triangulations present in the full spacetime geometries. Subsequently, in Sec. \ref{sec:phase-diagram} we provide further detail on the phase diagram of 3D CDT, and present the results of the numerical simulations that we used to study the effect of changing the spatial genus and triangulation ensembles on the structure of this phase diagram. Sec. \ref{sec:eff-act} contains a description of our investigation of the effective action for the volume profiles of the model, where we again vary both the spatial genus and the triangulation ensembles in order to check whether it leaves any imprint on the results. We summarize and discuss our overall findings in Sec. \ref{sec:discussion}. 

\section{Review of CDT}
\label{sec:spatial}
The framework of CDT is a non-perturbative approach to solve the Lorentzian quantum gravitational path integral \cite{ambjorn2012nonperturbative}, with the formal expression \eqref{eq:qg-pi}. This formal integral is implemented in CDT by introducing a regulator in the shape of a lattice cutoff, so that we perform a sum over triangulated geometries \footnote{Note that we also use the term `triangulation' for higher-dimensional simplicial manifolds throughout.} with a causal structure. Furthermore, there is a well-defined notion of Wick rotation in the model. This maps the geometries of Lorentzian signature to unique Riemannian geometries, and the sum over Lorentzian spacetimes with complex weights is transformed into a sum over states with real Boltzmann weights. This makes the ensemble amenable to sampling using Markov chain Monte Carlo methods that are typically employed in the context of statistical physics.

\begin{figure}[t]
	\centering
	\includegraphics[width=0.6\textwidth]{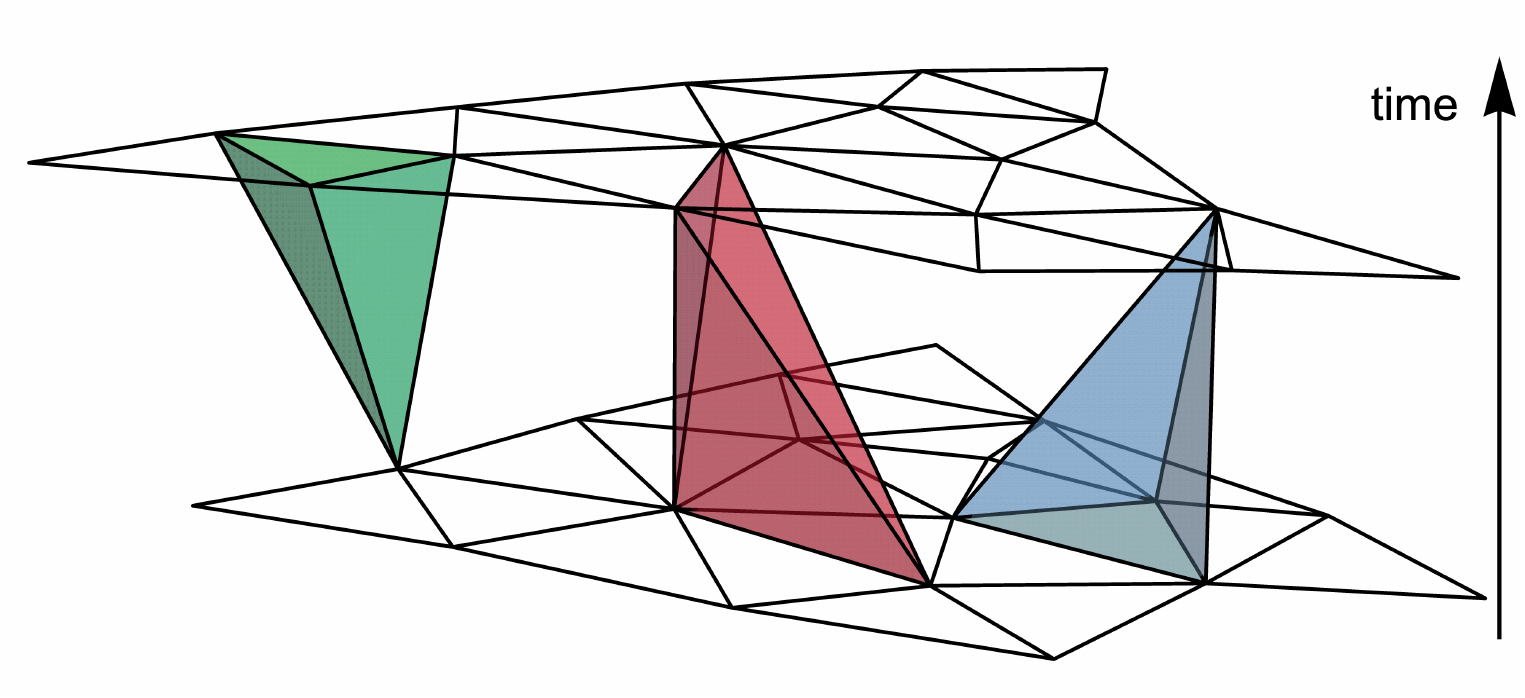}
	\caption{The three distinct types of building blocks in 3D CDT: a 13-simplex (left), a 22-simplex (center), and a 31-simplex (right). By convention, time labels increase in the upwards direction, so that the bottom slice is labeled $t$ and the top slice is labeled $t+1$.}
	\label{fig:simplices}
\end{figure}

The CDT geometries are simplicial manifolds with the added structure of a time foliation. The leaves of this foliation are spatial hypersurfaces of constant time. In what follows, we specialize to the case of CDT in (2+1) dimensions \cite{ambjorn2001lorentzian,ambjorn2001computer,ambjorn20023d}, which we simply refer to as 3D CDT, where the spatial hypersurfaces are two-dimensional triangulations. We assign a time label $t$ to each such slice, and neighboring slices labeled $t$ and $t+1$ are glued together with three-dimensional tetrahedra. We emphasize here that this time foliation does not correspond to a specific gauge choice, since there are no coordinates present in the system to start with. The edges connecting two distinct time slices are \emph{timelike} (and all of equal length $a_t$), whereas the edges confined to a single slice are \emph{spacelike} (also all of equal length $a_s$), with the relation $a_s^2 = -\alpha \cdot a_t^2$ between them, with $\alpha$ being a positive real number. We can then classify the 3D CDT simplicial building blocks as $(p,q)$-simplices, where a $(p,q)$-simplex has $p$ vertices in slice $t$ and $q$ vertices in slice $(t+1)$. The three possible simplex types are $(3,1)$, $(1,3)$, and $(2,2)$. We illustrate these three types of building blocks in Fig. \ref{fig:simplices}. For convenience, we abbreviate these types to 31-, 13-, and 22-simplices.

A consequence of using this basic set of building blocks is that the Boltzmann weights in the Wick-rotated CDT partition function take a simple form. The partition function of CDT after Wick-rotation is written
\begin{equation}
	\mathcal{Z} = \sum_{T \in \mathcal{T}} \frac{1}{C_T} e^{-S_\textrm{CDT} [T]},
	\label{eq:cdt-partition}
\end{equation}
where $C_T$ is the order of the discrete automorphism group of the triangulation $T$ and $S_\textrm{CDT}[T]$ is the discretized action obtained through the Regge prescription. For 3D CDT it takes the form
\begin{equation}
	S_\textrm{CDT}[T] = -k_0 N_0 + k_3 N_3,
	\label{eq:s-cdt}
\end{equation}
where $N_0$ is the number of vertices and $N_3$ the number of tetrahedra in the triangulation $T$. The coupling parameters $k_0$ and $k_3$ are related to the gravitational constant $G$ and the cosmological constant $\Lambda$, respectively. By definition, the individual type of simplices sum up to $N_3 = N_{31} + N_{13} + N_{22}$, where each of these terms counts the number of 31-, 13-, and 22-simplices, respectively. We furthermore have that $N_{31} = N_{13} = N_2^s$, where $N_2^s$ is the number of spatial triangles in the system.

We can now formulate the quantum averages of observables $\mathcal{O}$ in CDT as statistical expectation values:
\begin{equation}
	\langle \mathcal{O} \rangle = \frac{1}{\mathcal{Z}} \sum_{T \in \mathcal{T}} \frac{1}{C_T}\,\mathcal{O}[T]\,e^{-S_\textrm{CDT} [T]},
	\label{eq:q-exp-o}
\end{equation}
where $\mathcal{Z}$ is the partition function as defined in \eqref{eq:s-cdt}.

It is not known how to analytically compute the partition function \eqref{eq:cdt-partition} for 3D CDT. Therefore, we compute statistical estimates of the expectation value of an observable by sampling the CDT ensemble through Monte Carlo simulations.\footnote{Our implementation code can be found at \cite{brunekreef2022jorenb}.} We refer the interested reader to \cite{ambjorn2001computer} for more details on computer simulations of (2+1)-dimensional CDT.

For the purpose of this work, it was necessary to collect measurements for a given target system volume, and a standard method for achieving this is to introduce a quadratic volume-fixing term $S_\textrm{fix}$ in the bare action:
\begin{equation}
    S_\textrm{fix} = \epsilon \left( \tilde{N}_{31}-N_{31} \right)^2,
\label{eq:vol_fix}
\end{equation}
where $\tilde{N}_{31}$ is the target number of 31-simplices and $\epsilon$ is a constant that sets the strength of the volume-fixing. Including this term in the action causes $N_{31}$ to fluctuate around the target volume during the Monte Carlo simulations, and $\epsilon$ determines the typical size of the fluctuations. Note that we use $N_{31}$ for the fixing term instead of $N_3$, so that the system fluctuates around a target spatial volume as opposed to spacetime volume. This will turn out to be more convenient for our purposes, and it does not affect the phase structure of the system.

In the Introduction, we briefly mentioned some previously known results in 3D CDT that directly tie into the aim of this paper. We now summarize other earlier work that has served to shed more light on 3D CDT and its relation to continuum quantum gravity in three dimensions. It is notoriously difficult to obtain useful analytical results on the model, but several attempts have been made in this direction. The renormalization of the coupling constants in 3D CDT was investigated in \cite{ambjorn2004renormalization} using matrix model techniques. The authors of \cite{benedetti2007dimensional} determined the transfer matrix for the spatial slices of a reduced version of the model using similar techniques, which was subsequently used to derive a continuum Hamiltonian for this system. A different approach was followed in \cite{benedetti2017capturing}, where a balls-in-boxes model was used to describe the dynamics of the volumes of spatial triangulations in 3D CDT. 

Interesting results have also been obtained using numerical methods. The spectral dimension $d_s$ in the de Sitter phase of the model was studied in \cite{benedetti2009spectral}, where the measured results were compatible with the classical value of $d_s=3$ on large scales. Interestingly, the authors also found that the spectral dimension undergoes a dimensional reduction to a value compatible with $d_s=2$ on short scales, which is analogous to results obtained in four-dimensional CDT \cite{ambjorn2005spectral}. A generalized version of 3D CDT in which causality is preserved without a notion of preferred proper-time slicing was formulated and studied numerically in \cite{jordan2013causal,jordan2013sitter}, where it was found that the previously mentioned de Sitter volume profile still emerged in the simulations. The intrinsic structure of the spatial triangulations was investigated in \cite{brunekreef2022nature}, leading to the conjecture that the embedding of the slices in a three-dimensional geometry may give rise to a previously unknown model of two-dimensional quantum geometry. The relation of CDT to three-dimensional Ho\v{r}ava-Lifschitz gravity was explored in \cite{anderson2012quantizing}, and a numerical investigation of transition amplitudes between spatial triangulations was made in \cite{cooperman2014first}.

\section{Geometric ensembles}
\label{sec:ensembles}
In this section, we present a detailed discussion of the choice of ensemble $\mathcal{T}$ that is summed over in the partition function \eqref{eq:cdt-partition}. The standard CDT ensemble consists of simplicial manifolds, in which all $k$-simplices are uniquely identified by the $(k+1)$ vertices (or 0-simplices) comprising them. It is known that the simplicial manifold condition can be relaxed in two-dimensional Euclidean DT \cite{ambjorn1997quantumb} and two-dimensional CDT \cite{ambjorn2007putting} without altering the universal behavior in the continuum limit. No such universality result is known for CDT in dimensions three and higher. One of our objectives in this work is to investigate the consequences of relaxing the simplicial manifold conditions in 3D CDT. To this end, we define three ensembles that contain the standard CDT ensemble as a strict subset. These definitions are formulated as a set of constraints on the spatial triangulations of the system. Before discussing the ensembles in the context of 3D CDT, let us first define four classes of two-dimensional triangulations.

It is convenient to characterize these classes from the point of view of the 3-valent graphs dual to the triangulations. We only consider triangulations without boundary, so the dual graphs are vacuum Feynman diagrams of a $\phi^3$ scalar field theory. We follow the classification of such graphs as presented in \cite{ambjorn1997quantumb}. Allowing for all possible Feynman graphs of the $\phi^3$ theory gives us the so-called type I ensemble. The type II ensemble is obtained by excluding all graphs with tadpole insertions. If we furthermore exclude self-energy insertions, we find the type III ensemble. We also call members of this ensemble \emph{combinatorial} triangulations. Finally, we define an intermediate type IIA class of triangulations in which self-energy insertions are allowed, but only if each of these insertions contains two or more loops. Since triangulations of type I, II, and IIA do not satisfy the simplicial manifold conditions, we collectively refer to these classes as \emph{degenerate} triangulations.

\begin{figure}[htb]
	\centering
	\begin{subfigure}[t]{0.41\textwidth}
	\includegraphics[width=0.8\textwidth]{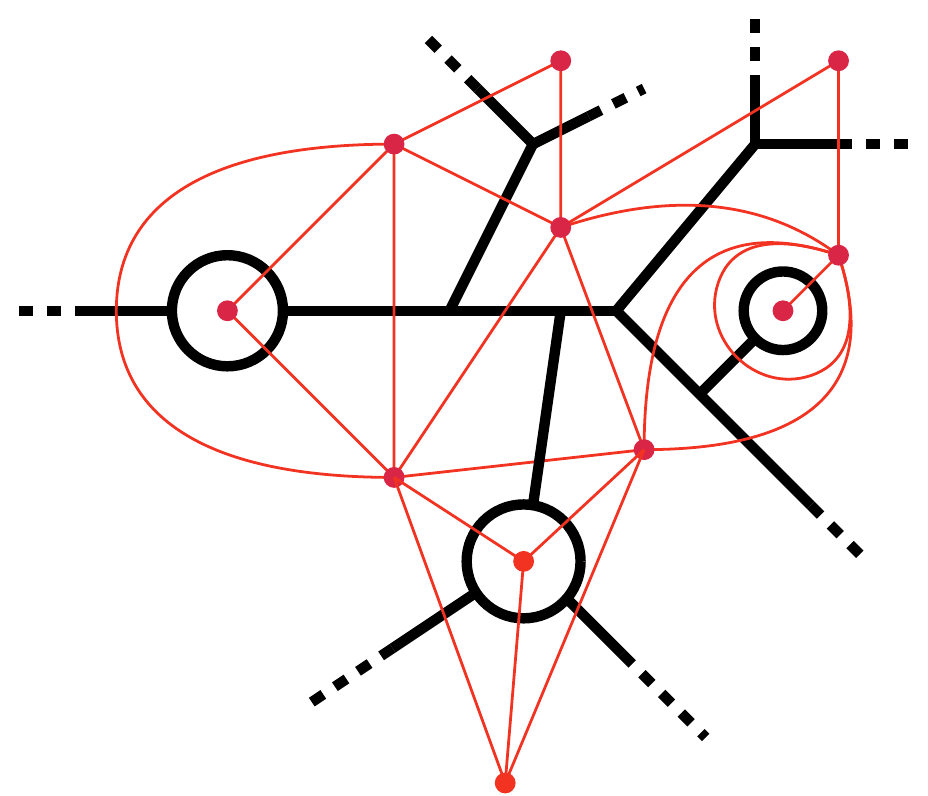}
	\caption{Type I}
	\label{fig:triangulation-type-0}
	\end{subfigure}
	\hfill
	\begin{subfigure}[t]{0.41\textwidth}
	\includegraphics[width=0.8\textwidth]{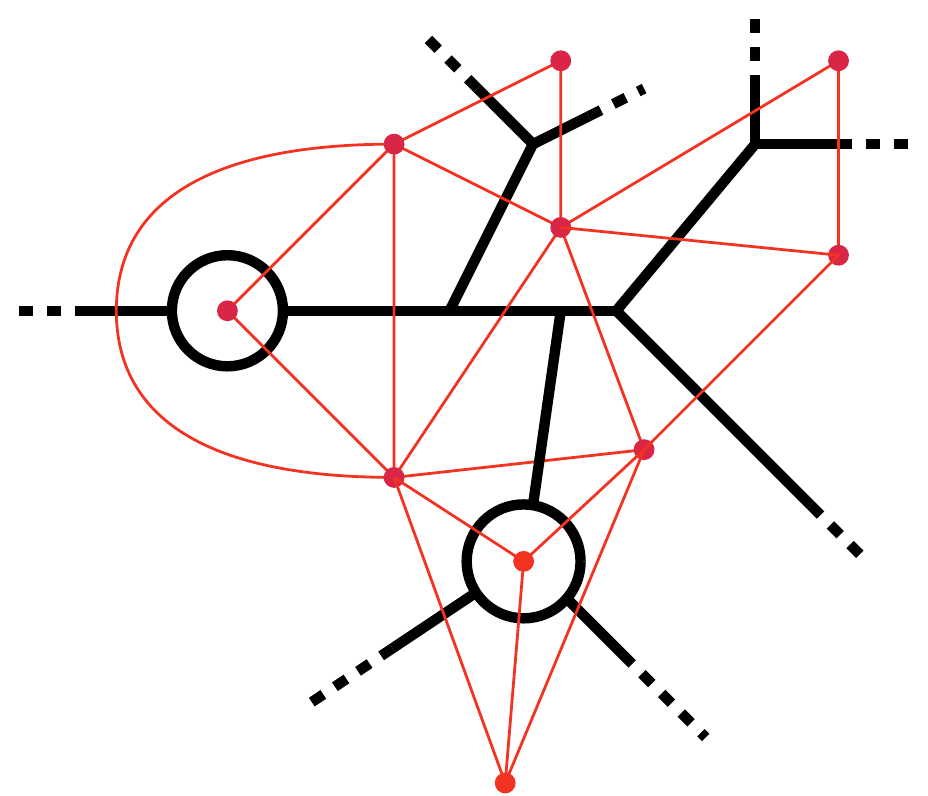}
	\caption{Type II}
	\label{fig:triangulation-type-1}
	\end{subfigure}
	\\
	\begin{subfigure}[t]{0.41\textwidth}
	\includegraphics[width=0.8\textwidth]{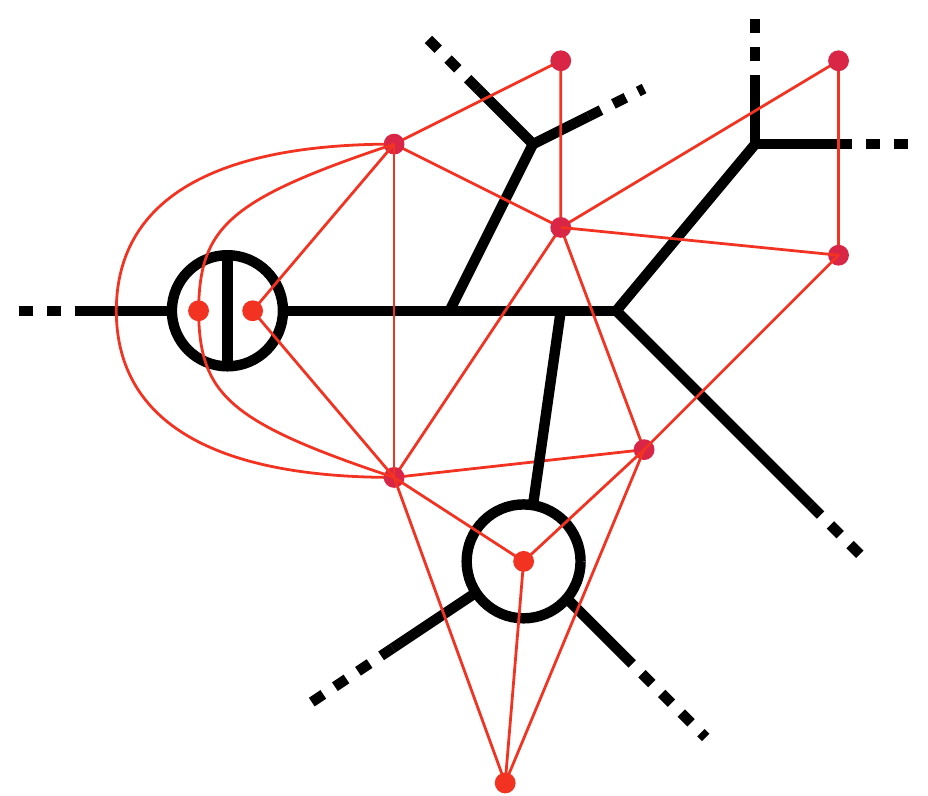}
	\caption{Type IIA}
	\label{fig:triangulation-type-2}
	\end{subfigure}
	\hfill
	\begin{subfigure}[t]{0.41\textwidth}
	\includegraphics[width=0.8\textwidth]{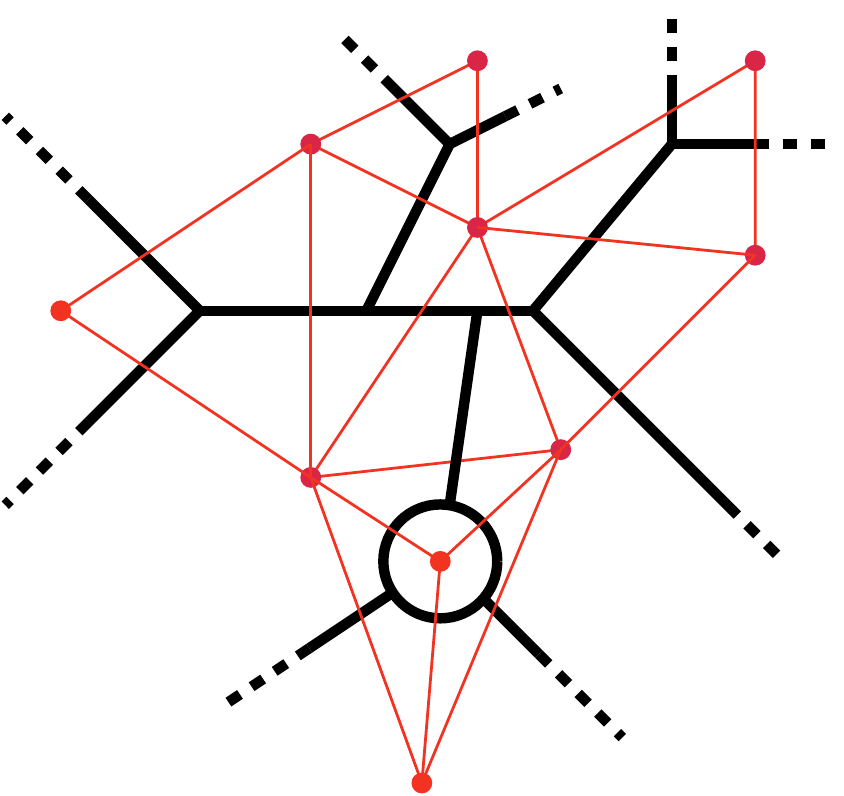}
	\caption{Type III}
	\label{fig:triangulation-type-3}
	\end{subfigure}
	\caption{Illustration of the four ensembles of spatial triangulations.}
	\label{fig:triangulation-types}
\end{figure}

We illustrate sample triangulation regions of the four types of ensembles in Fig. \ref{fig:triangulation-types}. The triangles of the triangulations (in red) dual to the $\phi^3$ graph (in black) should be considered equilateral, even though the embedding generally does not allow us to draw them with equal area, or even with straight edges. A type I triangulation (top left) are unrestricted, and the triangle surrounding the tadpole of the dual graph consists of only two vertices. This specific singular behavior is forbidden in a type II triangulation (top right), although the one-loop self-energy insertion produces a pair of triangles comprised of the same three vertices. A type IIA triangulation (bottom left) has no such pairs of triangles, although the two red links surrounding the dual graph two-loop self-energy insertion have the same two vertices, again violating the simplicial manifold condition. Finally, we show a region of a type III triangulation (bottom right) where such higher-loop self-energies are also forbidden in the dual graph, ensuring that the triangulation is a two-dimensional simplicial manifold.
 
From these four classes of two-dimensional triangulations, we can then define four ensembles of CDT triangulations by taking one of the classes as the possible choices for the spatial triangulations. What we then call the ``type III ensemble'' therefore corresponds to the standard CDT ensemble. Note that although the type I ensemble is the largest on our list, this is not the most general one we can formulate. One can relax further constraints on the ``slices'' at half-integer time where the 22-simplices appear as rectangles, but this is beyond the scope of this work.

It is technically challenging to implement the type I CDT ensemble in the current version of our simulation code. In this work, we therefore restrict our attention to the type II, type IIA, and type III ensembles. An important feature of CDT is that the topology of spatial slices is constant. This prevents the formation of baby universes that one encounters in Euclidean DT. Previous work on 3D CDT has focused on systems where the spatial slices are homeomorphic to a manifold $M$ of either spherical ($S^2$) or torus ($\mathbb{T}^2$) topology, i.e. surfaces of genus 0 and 1, respectively. The full space-time topology is then $M \times I$ for the case of open boundaries, and $M \times S^1$ for the case of periodic boundary conditions. In this work, we extend our investigation to ensembles where the spatial slices have genus $g > 1$. We constructed an algorithm for generating CDT configurations with $T$ time slices (with $T \geq 3$), where each time slice is a triangulation of a surface of arbitrary genus $g$, where $g$ is a nonnegative integer. The genus $g$ triangulations were constructed by first triangulating the surface of a $g$-holed torus $\mathbb{T}^g$, and subsequently gluing $T$ such slices together with interpolating 3-simplices. We finally impose periodic boundary conditions in the time direction for convenience, so that the resulting space-time manifold has topology $\mathbb{T}^g \times S^1$. Such a triangulation can subsequently be used as the initial configuration for Monte Carlo simulations, allowing us to sample the CDT partition sum for any spatial genus.

\section{Phase structure}
\label{sec:phase-diagram}
Simulations of 3D CDT have demonstrated the existence of two distinct geometric phases \cite{ambjorn2001nonperturbative}. The phase diagram of the model is spanned by the gravitational coupling $k_0$. The cosmological coupling $k_3$ is typically tuned to its $k_0$-dependent pseudocritical value $\kcc(k_0)$ in order to approach the limit of infinite volume. There is a first-order geometric phase transition at a certain critical value $\kgc$ of the gravitational coupling. This separates the phase space into two regions, and we will discuss the geometrical nature of these two phases in more detail in Sec. \ref{sec:eff-act}. In what follows, we present the results of our measurements of the phase diagram of 3D CDT for the distinct ensemble types and several values of the spatial genus $g$.

\subsection{Pseudocritical line for the cosmological coupling}
As a first step, we determined the shape of the pseudocritical line $\kcc(k_0)$ for the three distinct types of ensembles. Furthermore, we investigated the effect of the spatial genus on this pseudocritical line. In order to find $\kcc(k_0)$ at fixed $k_0$ for a certain choice of ensemble and spatial genus, we ran Monte Carlo simulations of such a system with target volume $\tilde{N}_{31}=10k$ and $T=3$ time slices, tuning the parameter $k_3$ until we reach a situation where $N_{31}$ fluctuates symmetrically around the target $\tilde{N}_{31}$. The number of time slices $T=3$ is the smallest allowed by our simulation code, so that the volume per slice is maximized. Investigations of the \emph{transfer matrix} \cite{ambjorn2013transfer} in (3+1)-dimensional CDT have indicated that using a setup like this is appropriate in that case. As an extra check, we have determined $\kcc(k_0)$ for a few fixed choices of $k_0$ using the same method, but with larger number of time slices up to $T=32$. Each time we found identical pseudocritical values $\kcc(k_0)$ when compared to the $T=3$ case.

\begin{figure}[ht!]
\centering
\includegraphics[width=0.75\textwidth]{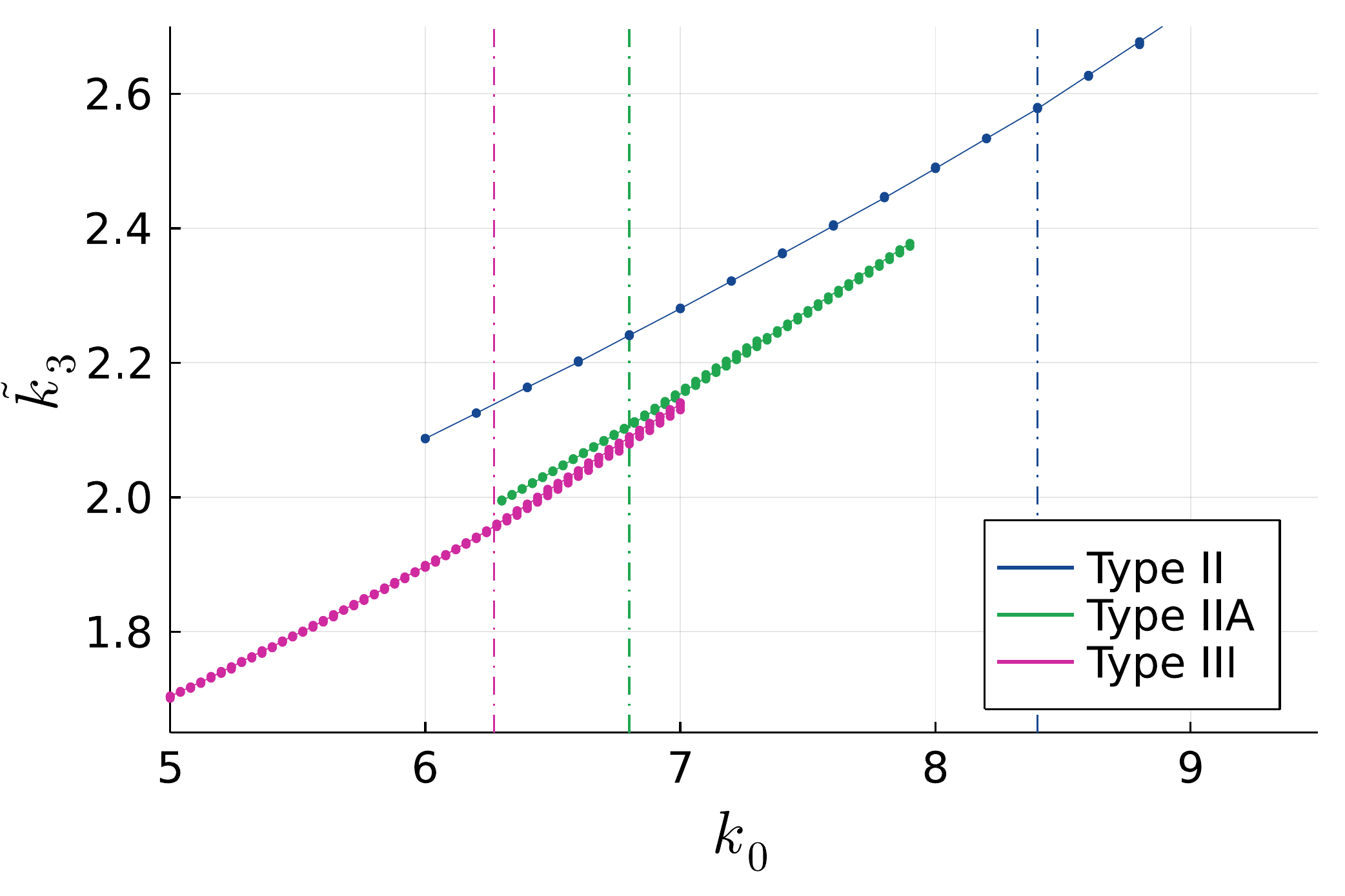}
\caption{The critical cosmological parameter $\kcc$ as a function of the bare gravitational parameter $k_0$ for the type II (blue), IIA (green), and III (purple) CDT ensembles with $T=3$ time slices and target volume $\tilde{N}_{31}=10k$. For the spatial genus, we used the values $g=0,1,2,15,30$. The thin lines are interpolations between the data points. The results are identical within measurement error (error bars are on the order of the dot size and therefore not shown) for different genera, making all the curves for a given ensemble overlap. The vertical dash-dotted lines indicate the approximate location of the phase transition. 
}
\label{fig:k3}
\end{figure}

The results of our measurements of the pseudocritical line are shown in Fig. \ref{fig:k3}. The ranges of $k_0$ for the different ensembles were chosen to be centered around their specific values of the critical $\kgc$. We explain how to find this critical point in the next part of this section. It is readily apparent that the choice of ensemble influences the shape of the critical line and the location of the critical point. This was to be expected, since the entropy of triangulations at fixed $\tn$ is different for each of the ensembles. Interestingly, the pseudocritical lines for the type IIA and type III ensembles are close together, while judging from Fig. \ref{fig:triangulation-types} one may have expected that the type II and type IIA systems do not differ much. The second interesting observation is that the pseudocritical lines are not affected by the choice of spatial genus. 

\subsection{Pseudocritical point for the gravitational coupling}
In order to determine the pseudocritical value $\kgc$ of the gravitational coupling we make use of the order parameter
\begin{equation}
    \optwo = \frac{N_{22}}{N_{31}},
    \label{eq:optwo}
\end{equation}
where $N_{22},N_{31}$ are the number of 22-simplices and 31-simplices, respectively. Note that $N_{31}$ equals the total number of spatial triangles in the geometry, whereas the 22-simplices function as connections between neighboring slices. Therefore, a larger value for $\optwo$ implies on average a higher level of interconnectivity between slices. The chosen order parameter $\optwo$ is conjugate to the parameter $k_0$ in the action \eqref{eq:s-cdt}. 

\begin{figure}[ht!]
\centering
\begin{subfigure}[t]{0.48\textwidth}
\includegraphics[width = 0.95\textwidth]{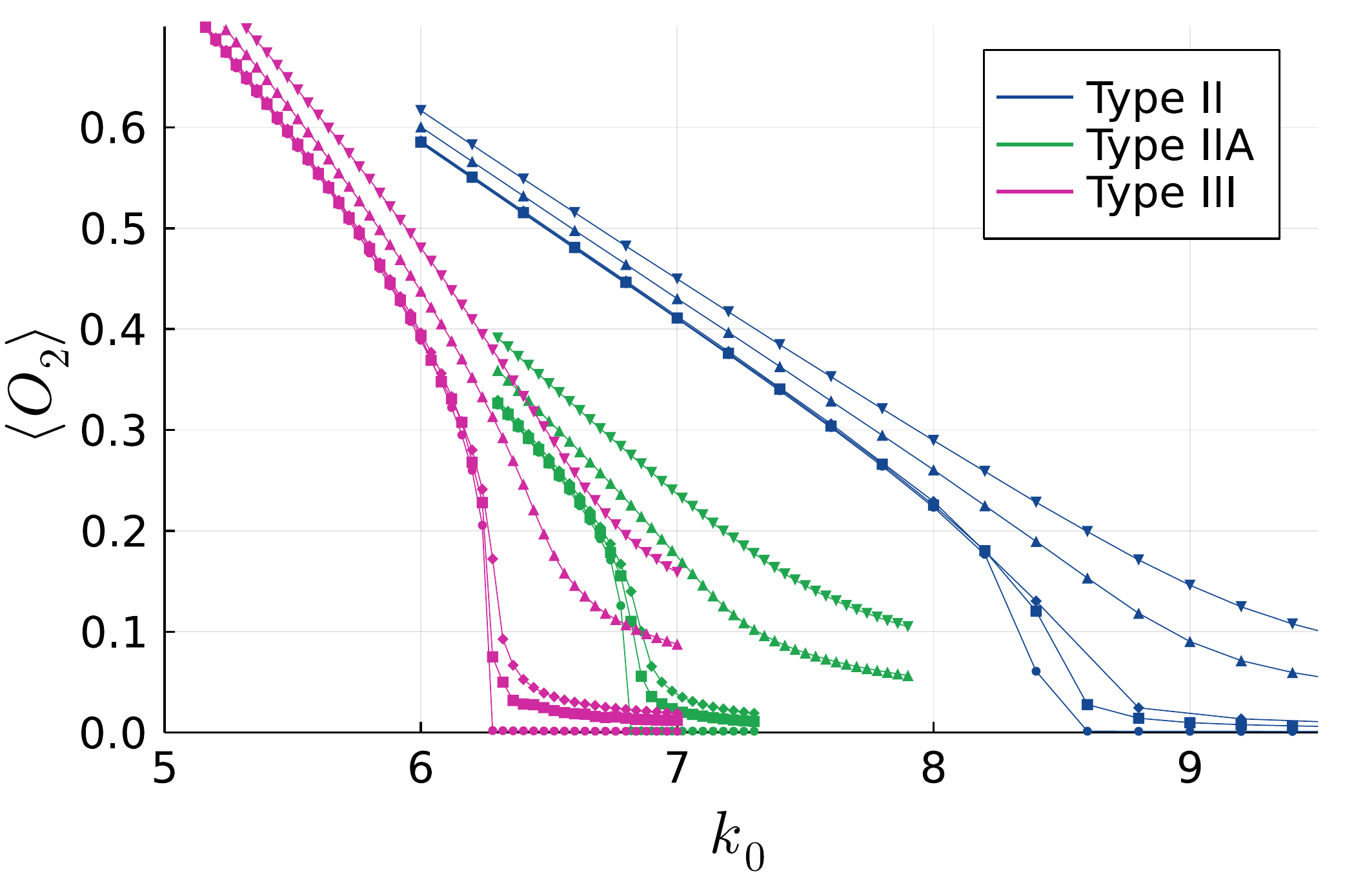}
\end{subfigure}
\hfill
\begin{subfigure}[t]{0.48\textwidth}
\includegraphics[width = 0.95\textwidth]{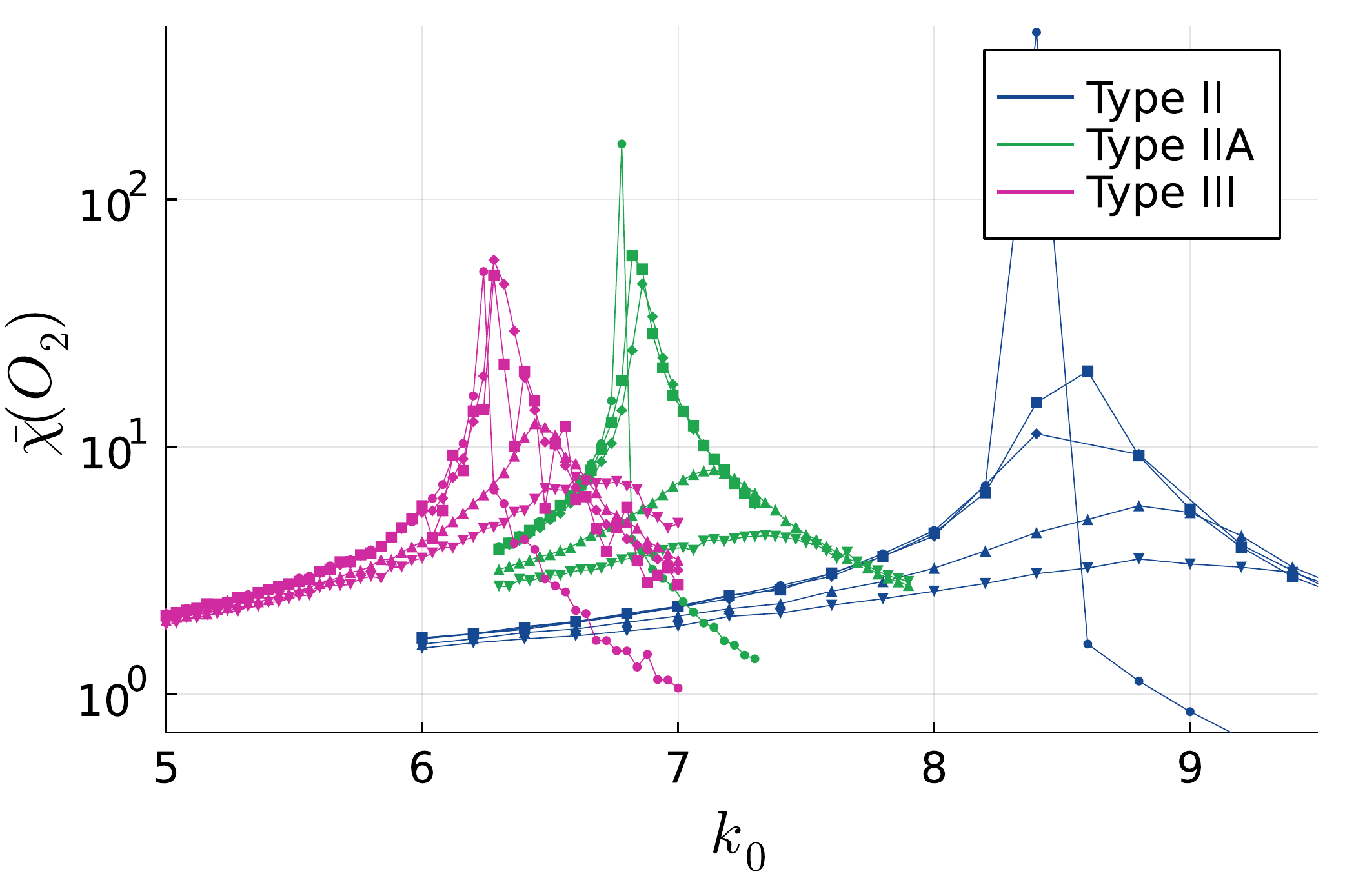}
\end{subfigure}
\caption{Left: the expectation value of the order parameter $\langle\optwo\rangle$ as a function of $k_0$. Right: 
the normalized susceptibility $\bar{\chi}(\optwo)$ as a function of $k_0$. The color coding refers to the type II (blue), IIA (green), and III (purple) of the ensembles. The distinct marker shapes correspond to genus $g= 0\,(\bullet) ,1\,(\sqbullet),2\,(\blackdiamond),15\,(\blacktriangleup),30\,(\blacktriangledown)$. The thin lines interpolate between the data points as a guide to the eye. Error bars are smaller than dot size, except for the susceptibilities near the peaks, where they grow larger, but we omit them for clarity.}
\label{fig:op2}
\end{figure}

We measured the ensemble averages of the order parameter $\optwo(k_0)$ using Monte Carlo simulation for the three types of ensembles, and for spatial genus $g \in \{0,1,2,15,30\}$. The systems we used in our simulations had $T=3$ time slices and target volume $\tn=10k$. The results are shown in Fig. \ref{fig:op2} (left). A similar structure appears for all three types of ensembles. As $k_0$ increases, the fraction of 22-simplices decreases. For small $g$, there is a sharp drop around a certain $k_0$, where the number of 22-simplices approaches zero. This can be interpreted as a geometric phase transition, and we provide further detail about the behavior of the two phases of the system in Sec. \ref{sec:eff-act} below. The drop softens for larger $g$ and further investigations have shown that this is a discretization artefact: increasing the system size of the simulations at high genus leads to sharper fall-offs, leading us to conjecture that a discontinuous drop is recovered in the infinite-volume limit. A related observation is the fact that the asymptotic minimum of $\optwo$ for the higher genus results tends to be higher than for small spatial genus, where it approaches zero. The reason for this is that a larger minimal number of 22-simplices is required to glue neighboring spatial slices of higher genus together. We therefore have a certain minimal value for the numerator in \eqref{eq:optwo}. Simulating larger systems for $k_0$ in this regime would keep the numerator approximately fixed to this minimal value, while increasing the denominator, so that the asymptotic value can be brought arbitrarily close to zero. As a final observation, we see that the location of the critical point differs for the three distinct choices of ensemble. As discussed before, this is entirely in line with expectations, and is a consequence of the entropy of configurations at a given target $\tn$.

We can estimate the location of the phase transition in the infinite-volume limit by measuring the susceptibility $\chi$ associated to the order parameter $\optwo$, which we define as
\begin{equation}
    \chi(\optwo) = N_{31} \cdot \left(\langle \optwo^2 \rangle - \langle \optwo \rangle^2\right).
\end{equation}
A peak in the susceptibility indicates the occurrence of a phase transition. We plot the normalized susceptibility $\bar{\chi}(\optwo) \equiv \frac{\chi(\optwo)}{\langle \optwo \rangle}$ in Fig. \ref{fig:op2} (right), measured for target volume $\tn = 10k$ and $T=3$ time slices. The peaks are wider for the ensembles of higher spatial genus, again implying that the precise location of the transition is harder to determine at the current system size. For each individual ensemble type, however, the transition points for different genera are close together, with the higher genus transition points shifted slightly to the right.

\subsection{Transition order}
The phase transition of standard 3D CDT with the type III ensemble and spatial genus zero is of first order \cite{ambjorn2001nonperturbative}. We investigated whether the order of the transition is influenced by the choice of ensemble and the genus of the spatial slices. Determining the order of the transition requires simulations of large systems for several values of the bare gravitational coupling $k_0$ close to the pseudocritical point $\kgc$, making the process very time-consuming. We therefore restricted our attention to the type II ensemble, which corresponds to the most relaxed set of conditions on the spatial triangulations that we considered in this work. Furthermore, we fix the spatial genus to that of a torus ($g=1$). 

\begin{figure}[ht!]
\centering
\begin{subfigure}[t]{0.48\textwidth}
\includegraphics[width = 0.95\textwidth]{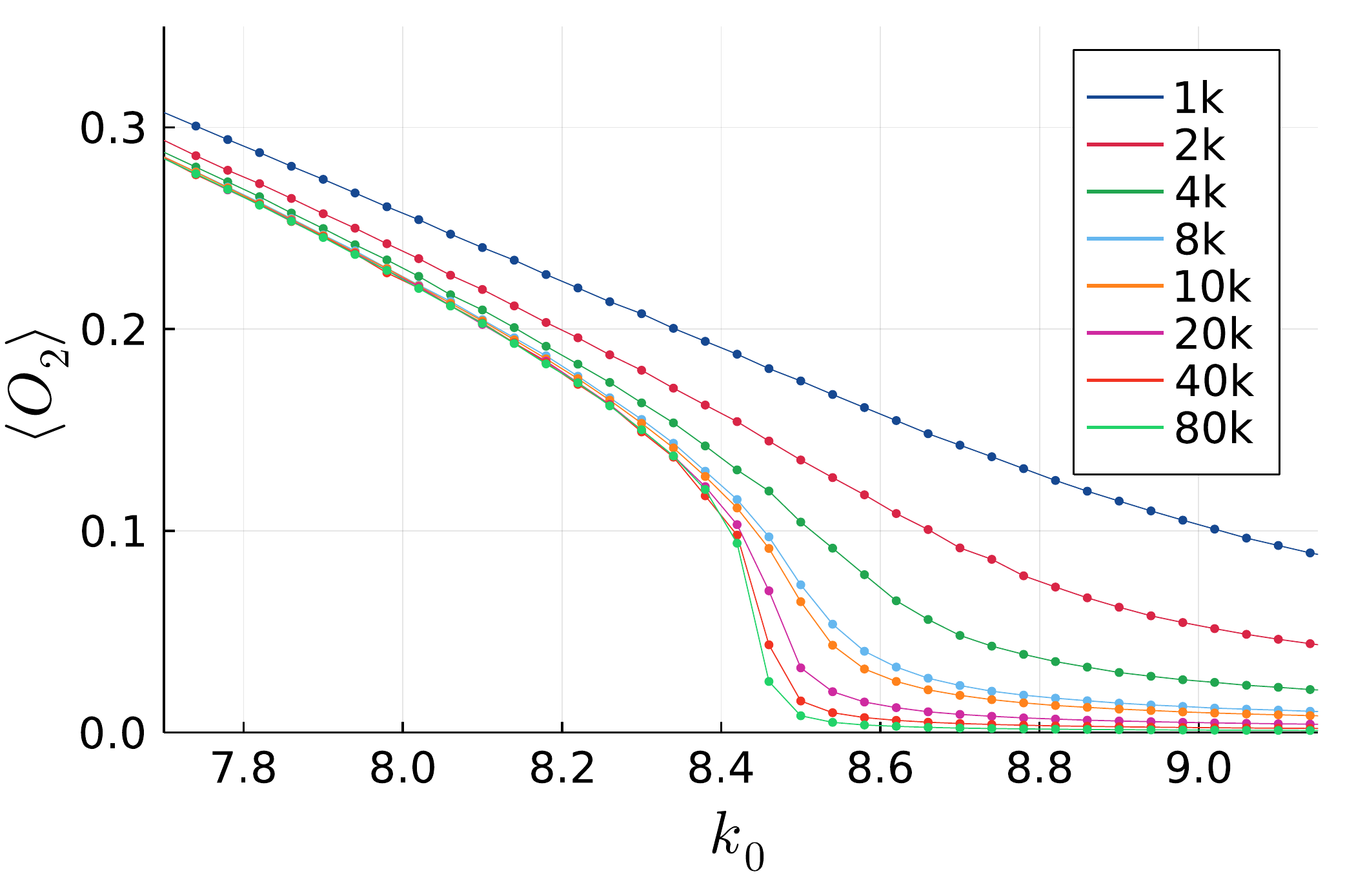}
\end{subfigure}
\hfill
\begin{subfigure}[t]{0.48\textwidth}
\includegraphics[width = 0.95\textwidth]{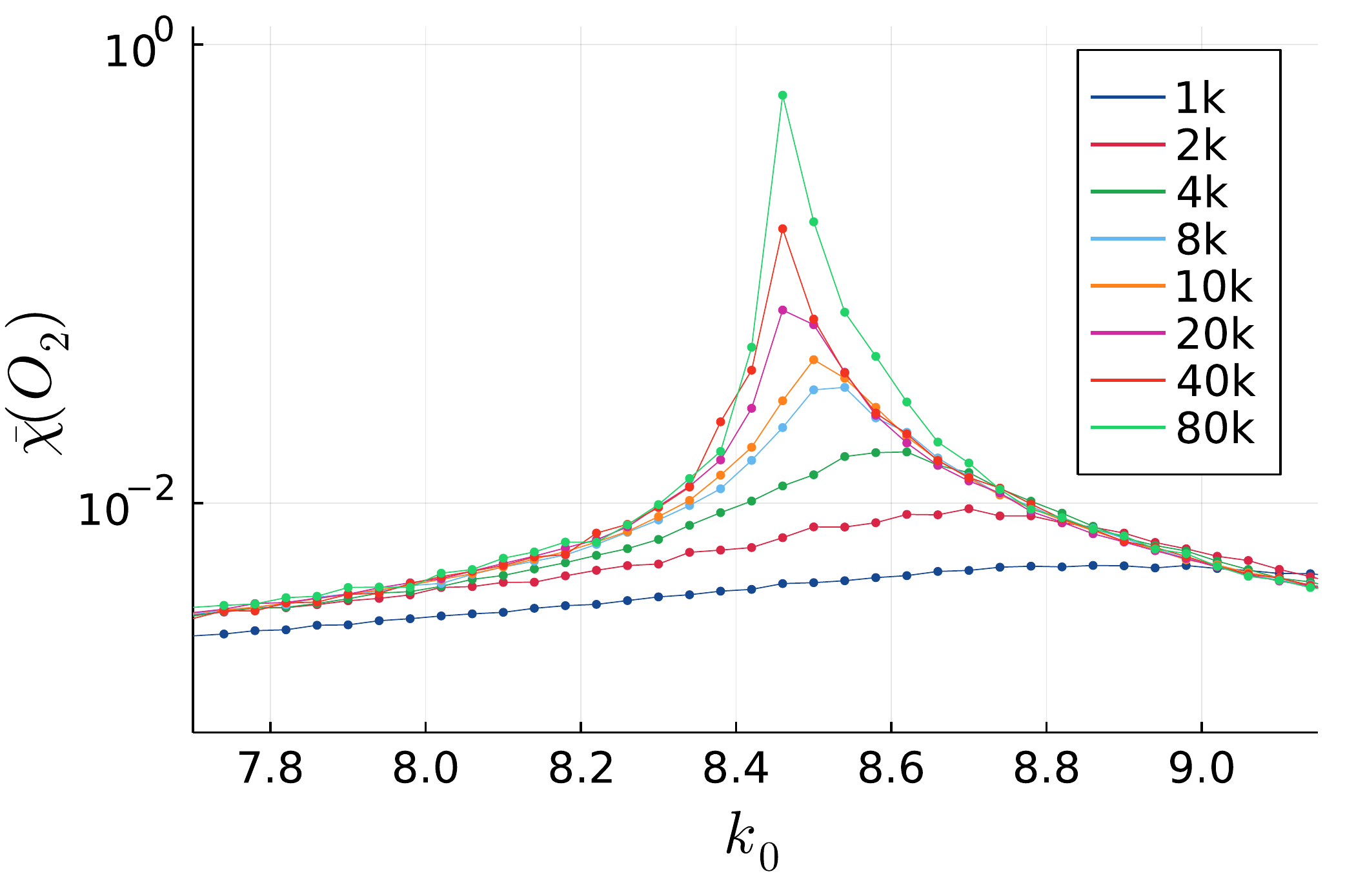}
\end{subfigure}
\\
\begin{subfigure}[t]{0.48\textwidth}
\includegraphics[width = 0.95\textwidth]{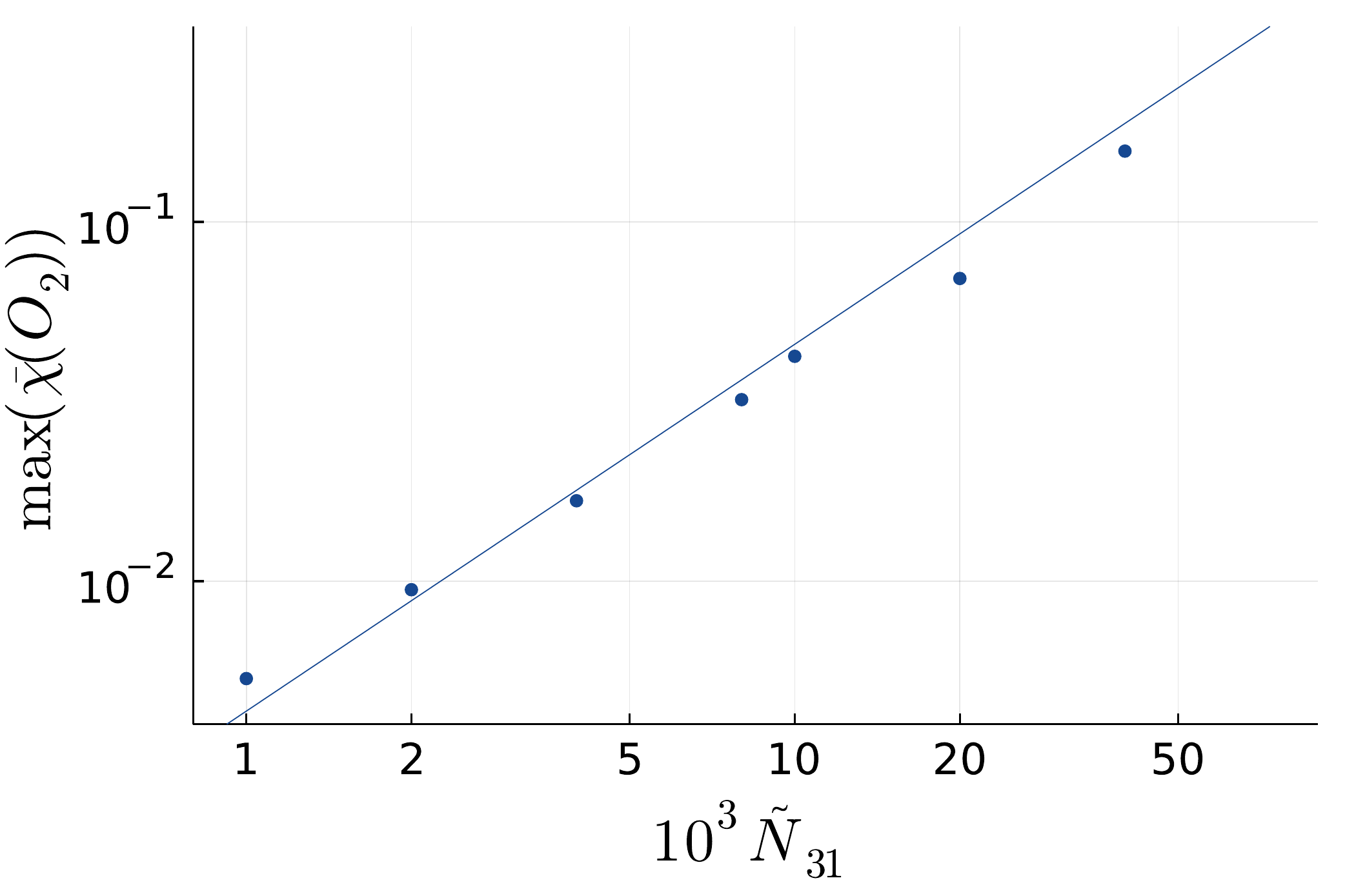}
\end{subfigure}
\hfill
\begin{subfigure}[t]{0.48\textwidth}
\includegraphics[width = 0.95\textwidth]{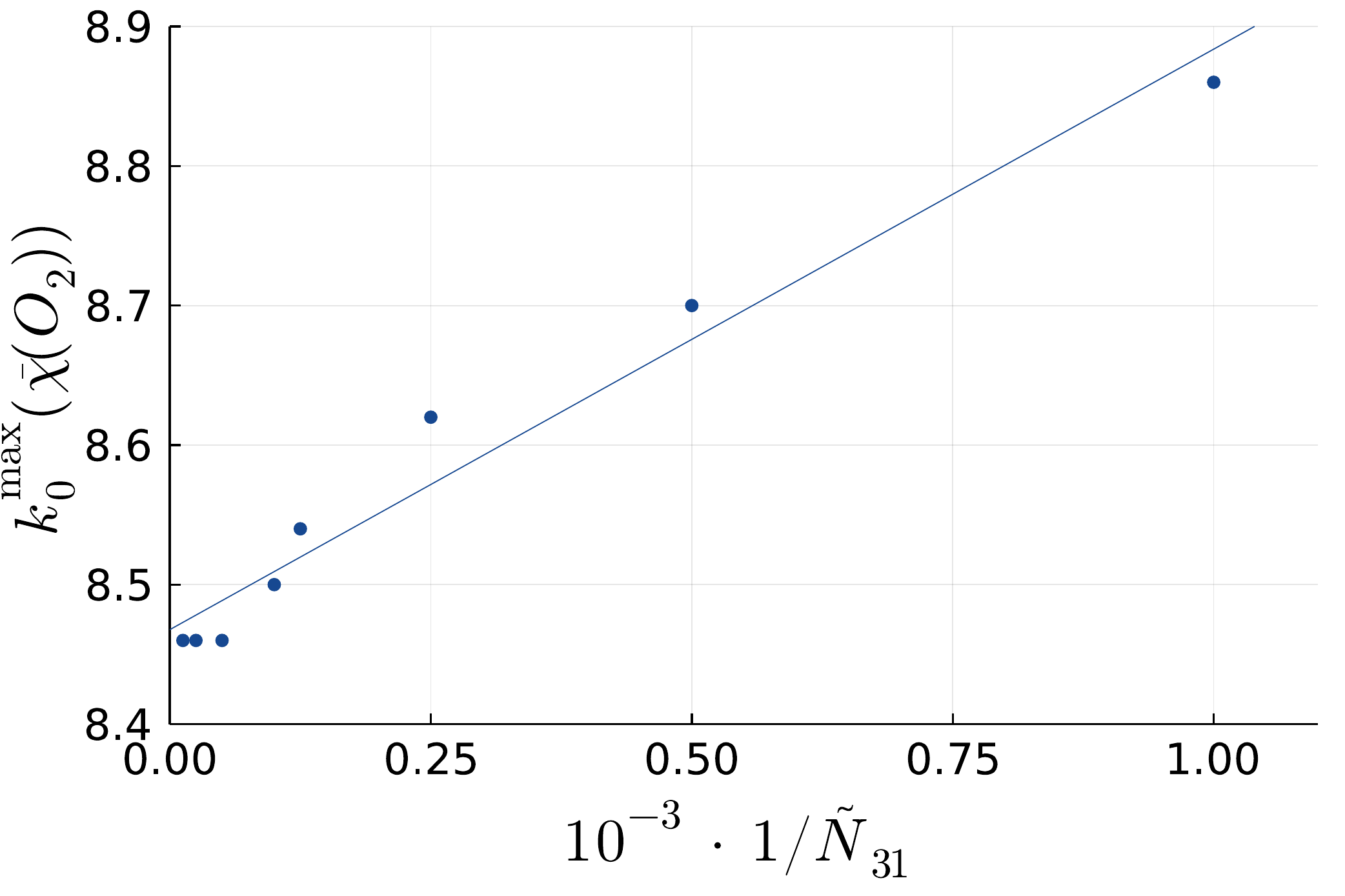}
\end{subfigure}
\caption{Top left: the expectation value of the order parameter $\langle \optwo \rangle$ as a function of $k_0$ for nine distinct system volumes with spatial genus $g=1$. Top right: the susceptibility $\bar{\chi}(\optwo)$ as a function of $k_0$. Bottom left: the height of the peaks of the normalized susceptibility as a function of the target system volume $\tn$. Bottom right: the location of the peaks of the susceptibility as a function of the inverse target system volume $1/\tn$. Error bars for the plots in the top row are smaller than dot size, except near the peaks of the susceptibility, where they are also omitted for clarity.}
\label{fig:v-op2}
\end{figure}

Previously, we showed how the transition manifests itself through the order parameter $\optwo$. In fact, true phase transitions cannot exist in finite systems like the ones we simulate \cite{goldenfeld2019lectures}, although their presence can be inferred by performing a scaling analysis of the behavior of the order parameter near the supposed critical point. In Fig. \ref{fig:v-op2} we present such a scaling analysis based on the ensemble average and fluctuations of the order parameter $\optwo$ for nine distinct target system volumes $\tn$ in the range $ \left[1k,80k\right]$. We show a plot of the expectation value $\langle \optwo\rangle$ in the top left of Fig. \ref{fig:v-op2}, from which it is readily apparent that the jump to approximately zero becomes more pronounced as the volume grows larger, which is a strong hint of a first-order phase transition in the infinite-volume limit.

The normalized susceptibility $\bar{\chi}(\optwo)$ is plotted in the top right of Fig. \ref{fig:v-op2}. If the transition is first-order, we expect the susceptibility to diverge like $\sim \tn$ at the critical point as the volume approaches infinity. Furthermore, the location of the peak is expected to shift like $\sim \tn^{-1}$. It is clear from the results that the expected peak emerges, the maximum grows with increasing volume, and its location shifts leftwards. The heights $\textrm{max}(\bar{\chi}(\optwo))$ (bottom left) and locations $k_0^\textrm{max}(\bar{\chi}(\optwo))$  (bottom right) are plotted as a function of the (inverse of the) volume. The straight line in the bottom left plot (showing the peak heights) is a fit of the form $c \cdot \tn^q$, where we find a best fit exponent $q = 1.04(6)$. This is consistent with the aforementioned $\sim \tn$ divergence of the peak height. The straight line in the bottom right plot (showing the location of the peaks) is a linear fit of the form $a+c \cdot \tn^{-1}$, also showing that the location of the peak shifts like $\sim \tn^{-1}$ as expected of a first-order transition. It should be pointed out that the largest volumes $\tn = 20k, 40k, 80k$ now appear near the origin of this plot, and that the fact that the peak locations $k_0^\textrm{max}$ are equal for these three volumes is an artefact of the spacing we used for the distinct $k_0$. In summary, we interpret these results as being consistent with a first-order transition in the type II ensemble with spatial genus 1. Further analysis also showed the emergence of a double-peak structure in the histogram of measurements of $\optwo$ taken very close to the pseudocritical point, with the peaks growing in relative size as the volume was increased. This is another clear sign characteristic of a first-order transition.

We conclude that the phase transition for the type II, spatial genus 1 ensemble is likely of first order. It is then reasonable to expect that the type II, spatial genus 0, and the type III, spatial genus 1 ensembles both have transitions of first-order nature as well. We performed exploratory simulations of these ensembles, and although we did not collect as large statistics as for the type II, genus 1 ensemble, all results are indeed consistent with a first-order transition in both cases. These results suggest that the phase transition of 3D CDT is of first order in general, and that changing the manifold constraints and the topology of the spatial slices do not modify its nature.

Since our code currently cannot handle the type I ensemble, we have not been able to investigate the phase diagram (and the expected critical point) there. Furthermore, it would be interesting to study the transition for higher values of the spatial genus. However, as we have seen earlier, we require much larger system sizes to observe a sharp transition for higher spatial genera, making an investigation of the critical point significantly more time-consuming. Addressing these issues is beyond the scope of this work, and we leave them open for possible future research.

\section{The effective action}
\label{sec:eff-act}
In this section, we discuss the behavior of spatial volumes in the phase $k_0 < \kgc$ (called the \emph{de Sitter} phase), and the determination of an \emph{effective action} that describes their ensemble average and typical fluctuation size. Such effective actions have been studied extensively in the context of (3+1)-dimensional CDT \cite{ambjorn2008planckian,ambjorn2008nonperturbative,ambjorn2013transfer,ambjorn2014effective,ambjorn2016impact,ambjorn2017fourdimensional}. In the (2+1)-dimensional setting, the effective action was investigated for spatial topologies of the sphere $(g=0)$ \cite{ambjorn20023d} and torus $(g=1)$ \cite{budd2013exploring}. We set out to extend this analysis to systems with higher spatial genus on the one hand, and on the other hand study the effect of relaxing the simplicial manifold constraints on the measured form of the effective action. Before proceeding to our results concerning the effective action, we briefly discuss the \emph{volume profile} observable in CDT. We summarize known results in the (2+1)-dimensional version of the model, where it was found that an effective volume action can indeed be formulated for the systems with spatial genus $g=0,1$ by measuring the covariance matrix of the volume profiles.  We subsequently present our measurements of the volume covariance matrices for several values of the spatial genus $g$, going beyond the cases of $g=0,1$, and furthermore extend the computer simulations to the ensembles with relaxed simplicial manifold constraints. Our results suggest that an effective action can also be formulated for spatial genus $g \geq 2$, and for all three ensemble types that we studied.

\subsection{Volume profiles}
A simple observable that is often used to study CDT ensembles is the volume profile $\langle n(t) \rangle$, where $n(t)$ is the spatial volume of the time slice with label $t$, and the angle brackets indicate the quantum expectation value as defined in \eqref{eq:q-exp-o}. In the context of 3D CDT, we define the spatial volume to be the number of spatial triangles contained in the slice.

\begin{figure}[htb]
	\centering
	\begin{subfigure}[t]{0.46\textwidth}
	\includegraphics[width=0.9\textwidth]{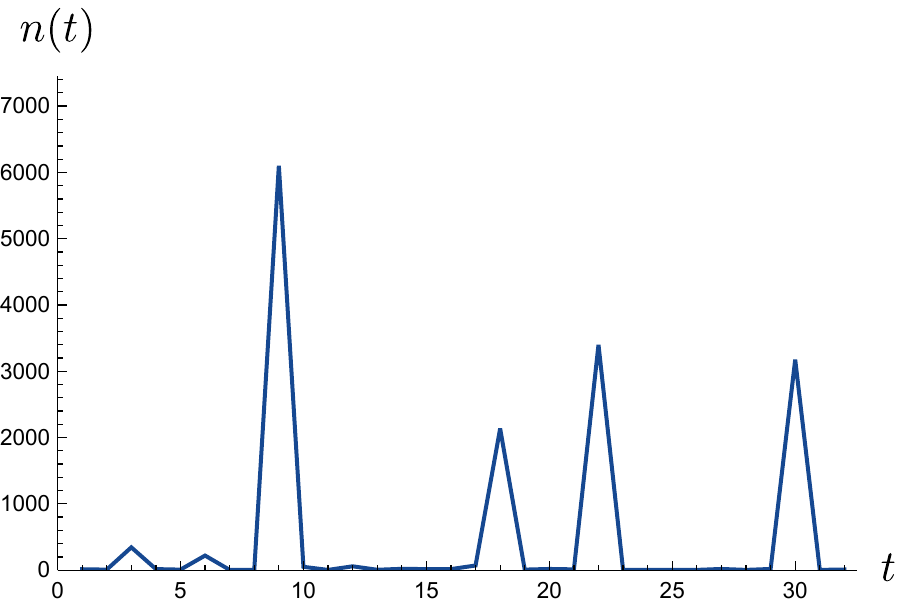}
	\end{subfigure}
	\hfill
	\begin{subfigure}[t]{0.46\textwidth}
	\includegraphics[width=0.9\textwidth]{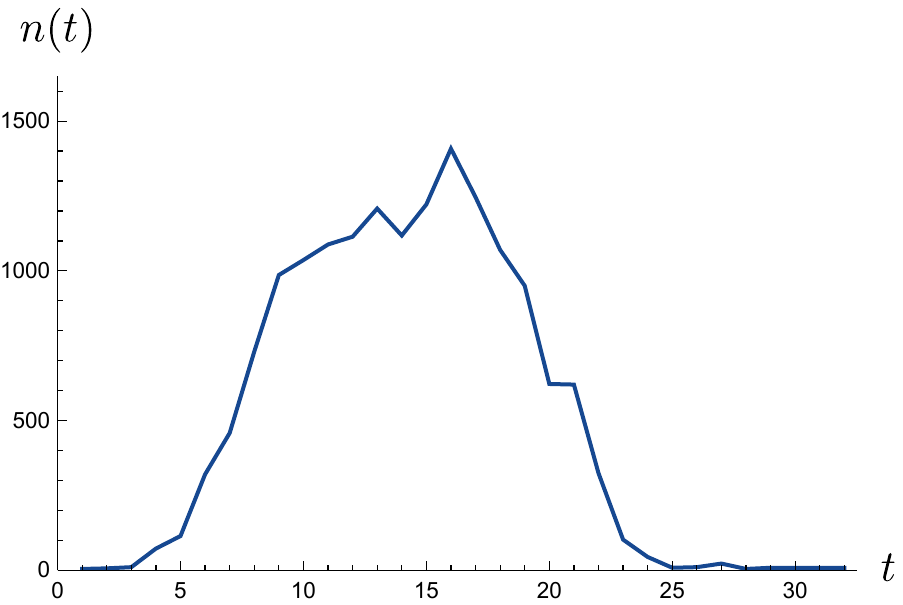}
	\end{subfigure}
	\par\bigskip
    \begin{subfigure}[t]{0.46\textwidth}
	\includegraphics[width=0.9\textwidth]{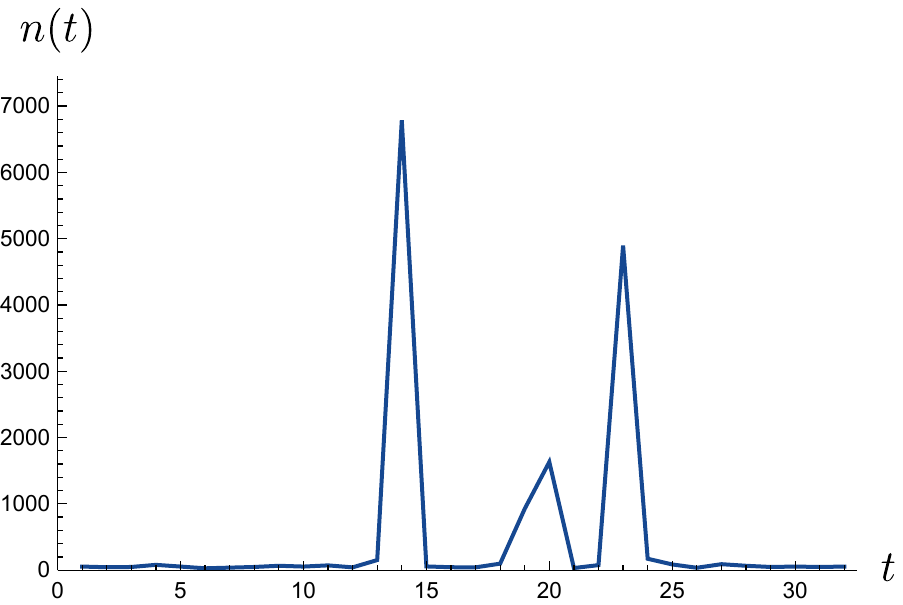}
	\end{subfigure}
	\hfill
    \begin{subfigure}[t]{0.46\textwidth}
	\includegraphics[width=0.9\textwidth]{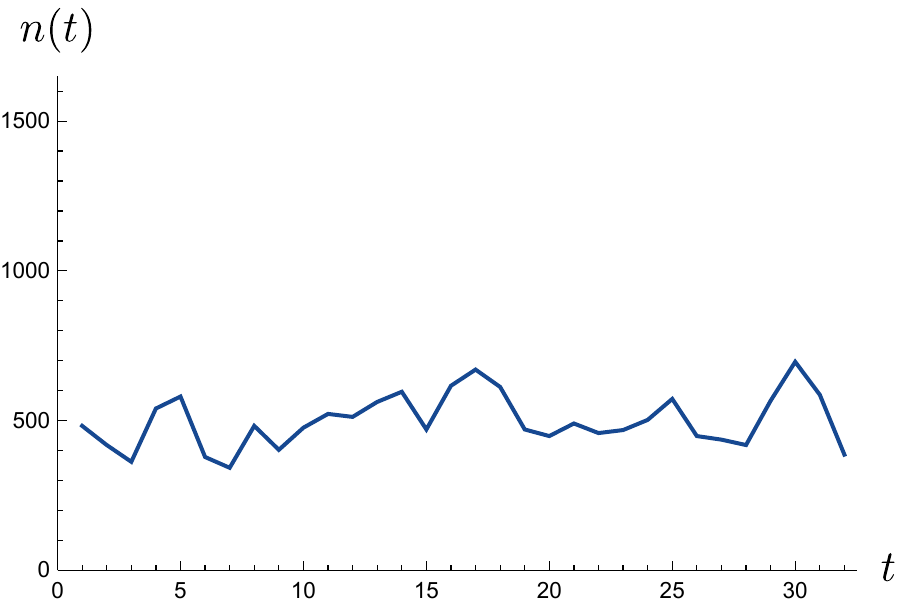}
	\end{subfigure}
	\caption{Snapshot volume profiles of geometries in the type III CDT ensemble, with $T=32$ time slices and target volume $\tn=16k$. The profiles on the left were obtained in the degenerate phase of the system, with $k_0 = 8.0$. The profiles on the right are taken from the de Sitter phase, with $k_0 = 5.0$. The top two profiles were taken from the ensemble with spatial genus 0, and the bottom two profiles at spatial genus 1.}
	\label{fig:snapshot-vps}
\end{figure}

Computer simulations have demonstrated \cite{ambjorn20023d,budd2013exploring} that the expectation value of the volume profile of 3D CDT manifestly differs between the two phases $k_0 < \kgc$ and $k_0 > \kgc$. For $k_0 > \kgc$, which we call the \emph{degenerate} phase, the volumes of the spatial slices are largely independent from one another, and only very minor correlations are observed between neighboring slices. The volume of the spatial slices can therefore jump significantly in just a single time step. For $k_0 < \kgc$, which we call the \emph{de Sitter} phase, the behavior is very different. In a typical geometry in this phase, volumes of neighboring slices are strongly correlated, resulting in a well-defined average `classical' background spacetime. Furthermore, we observe small quantum fluctuations around this background geometry. The origin of the name of this phase is the fact that the average background volume profile can be matched to that of a three-dimensional Euclidean de Sitter spacetime. We present snapshot volume profiles of individual CDT geometries appearing in our Monte Carlo simulations in Fig. \ref{fig:snapshot-vps}, for both the degenerate and de Sitter phases with spherical and toroidal spatial topology. The measurements were performed for the type III ensemble with target volume  $\tn = 16k$ and $T=32$ time slices.
We can distinguish three types of behavior in the snapshot profiles. The slice volumes of neighboring slices in the degenerate phase are largely uncorrelated, both for spherical $(g=0)$ and toroidal $(g=1)$ spatial topologies. However, in the de Sitter phase, we observe that a `blob' of finite time extent emerges for $g=0$, whereas the volume is spread out over the full time extent in the case of $g \geq 1$. Neighboring slices now tend to be of comparable volume. The fact that such an extended geometry emerges in the de Sitter phase suggests that it makes sense to compute its quantum average. Since the time labels $t$ have no intrinsic meaning, it is natural to distinguish two cases. Snapshots of the geometries with $g\geq 1$ exhibit a time translation symmetry, implying that the most relevant observable in this situation is the doubly averaged quantity
\begin{equation}
    \langle n \rangle = \frac{1}{T} \sum_t^T \langle n(t) \rangle.
\end{equation}
On the other hand, the individual geometries for $g=0$ exhibit a clustering of the volume in a finite region, resulting in the blob-like structure mentioned earlier. In order to find the average shape of this blob, we therefore shift all individual snapshot volume profiles of this ensemble to their center of mass (taking into account the periodicity) before computing the quantum average $\langle n(t) \rangle$. 

\begin{figure}[ht!]
\centering
\includegraphics[width = 0.45\textwidth]{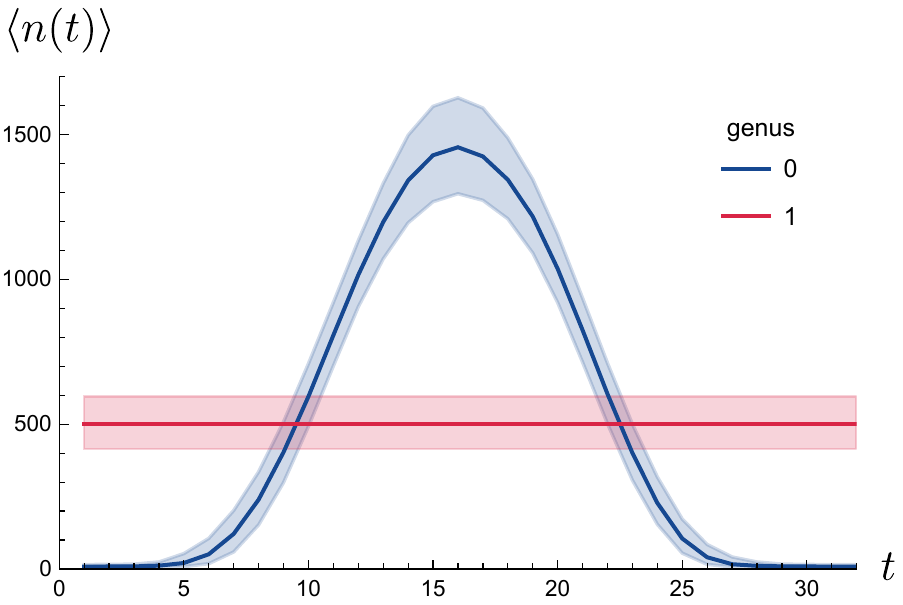}
\includegraphics[width = 0.45\textwidth]{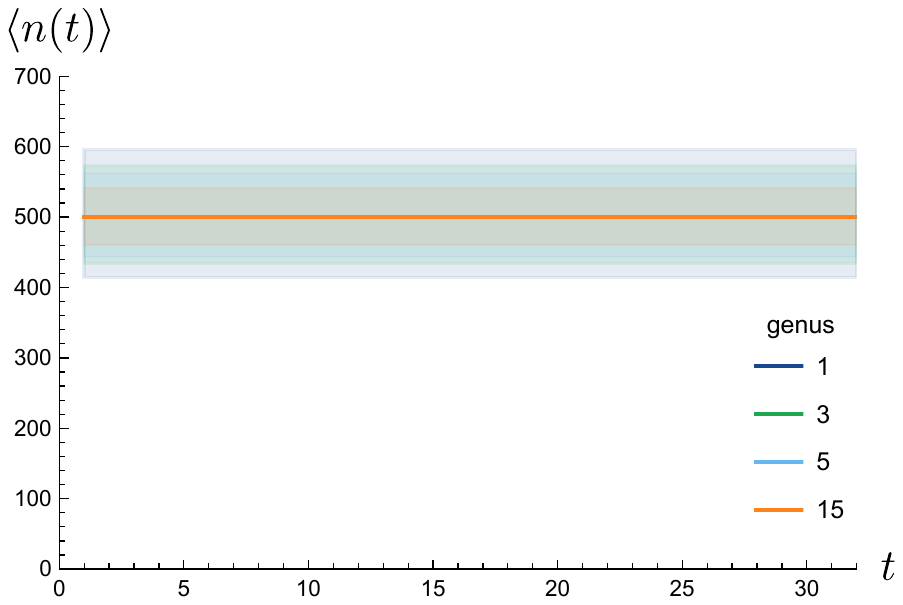}
\caption{Left: the average volume profiles for the type III CDT ensemble with spatial genus 0 (blue) and spatial genus 1 (red). Right: the average volume profiles for the type III CDT ensemble with spatial genus $g=1,3,5,15$, which are seen to coincide. The shaded regions indicate the typical size of the fluctuations around the mean, where the width of the band decreases with increasing genus.}
\label{fig:sphere-torus-vp}
\end{figure}

In order to improve our understanding of 3D CDT with spatial genus $g \geq 2$, we measured the average volume profiles in the de Sitter phase for the type III ensemble with spatial genus $g=0,1,3,5,15$ through Monte Carlo simulations with target volume $\tn=16k$ and $T=32$ time slices. We show the results of these measurements in Fig. \ref{fig:sphere-torus-vp}. The shaded region indicates the typical size of fluctuations around the average. Note that the $\langle n(t) \rangle$ used for the torus and higher genera is the constant value $\langle n \rangle$ as defined above.
Our measurements show that the expectation values of the higher-genus volume profiles are identical (within statistical error) to that of the system with spatial torus topology. Furthermore, the fluctuations around the average are small and approximately follow a Gaussian distribution. We interpret this as evidence that we can construct an effective action that describes the behavior of volume profiles, analogous to the findings from \cite{ambjorn20023d,budd2013exploring}.

\subsection{The effective action}
Suppose there exists an effective action $\seff[n_t]$ for the spatial volume $n_t := n(t)$ as a function of time label $t$, with a classical solution $\avn$ satisfying $\left.\delta \seff[n_t]\right|_{n_t = \avn}=0$. Quantum fluctuations around this classical solution can be described by expanding the effective action to quadratic order:
\begin{equation}
    \seff[\avn + \delta n_t] = \seff[\avn] + \frac{1}{2} \sum_{t, t'} \delta n_t P_{tt'}  \delta n_{t'} + \mathcal{O}(\delta n_t^3),
\end{equation}
where the linear term vanishes since it is evaluated at the classical solution and the operator $P_{tt'}$ is written
\begin{equation}
    P_{tt'} = \left.\frac{\partial^2 \seff}{\partial n_t \partial n_{t'}}\right|_{n_t = n_{t'} = \avn}.
    \label{eq:ptt}
\end{equation}
This approximation is valid as long as the fluctuations are Gaussian, which we indeed found to be the case in our aforementioned result. The covariance matrix $C_{tt'}$ of the quantum fluctuations is equal to the inverse of the operator $P_{tt'}$: 
\begin{equation}
    C_{tt'} = \langle \delta n_t \delta n_{t'} \rangle = \left(P^{-1}\right)_{tt'}.
\end{equation}
This allows us to construct the effective action for 3D CDT for a given spatial genus $g$ by measuring the covariance matrix using computer simulations. 

The way this works is as follows. We measure the ensemble averages of the covariance matrices for 3D CDT with spatial genus $g$ for a range of target system volumes $\tilde{N}_{31}$. We do not measure precisely \emph{at} the fixed target volumes, but rather let the system fluctuate \emph{around} a given $\tn$. The reason for this is that measuring at fixed volume introduces a zero mode in the covariance matrix that has to be projected out before computing the inverse, but this issue can be overcome by introducing the quadratic volume fixing term defined in \ref{eq:vol_fix}. Adding this term to the Regge action forces the volume to fluctuate around the target volume $\tn$ with typical fluctuation size determined by the parameter $\epsilon$. For our simulations, we set $\epsilon = 4 \cdot 10^{-5}$. It is important to note that the presence of a volume-fixing term also influences the measured effective action, however, it is straightforward to correct for this by subtracting
\begin{equation}
    \frac{\partial^2 S_\textrm{fix}}{\partial n_t \partial_{n_{t'}}} = 2 \epsilon
\end{equation}
from the measured matrix $P_{tt'} = (C^{-1})_{tt'}$.

The systems we use for our measurements have $T=4$ time slices. This choice was made since it maximizes the volume per slice while still explicitly demonstrating the volume-fixing correction to the inverse covariance matrix elements $P_{02}, P_{20}$. We ran tests with systems of $T=20$ time slices in the initial stage of our research, where we compared the resulting covariance matrices with the $T=4$ case, finding that the latter is indeed an appropriate substitute. A further advantage of using $T=4$ is that the `blob' structure shown in Fig. \ref{fig:sphere-torus-vp} for spatial genus 0 does not emerge for systems of such small time extent. This makes the system translationally invariant just as for the higher-genus cases.

It is clear from \eqref{eq:ptt} that measuring the covariance matrices only provides us with information on the second derivatives of the effective action with respect to the average slice volume. The reason we probe the system at a range of distinct target volumes is that we want to reconstruct the effective action from its second derivatives, although, this still requires us to formulate an ansatz for the functional form of the effective action to which we can subsequently match the results. It is \emph{a priori} unclear what this ansatz should be. However, building upon earlier work in 3D \cite{ambjorn20023d,budd2012effective,budd2013exploring} and 4D \cite{ambjorn2008planckian,ambjorn2008nonperturbative,ambjorn2013transfer,ambjorn2014effective,ambjorn2016impact,ambjorn2017fourdimensional}, we assume the presence of a \emph{kinetic} term that is quasilocal in time, and a \emph{potential} term that is local in time. The cited work takes inspiration from a minisuperspace treatment of the continuum quantum gravitational path integral, which in 3D results in the following effective action for the spatial volume $V(t)$:
\begin{equation}
\label{eq:minisuper}
    S_{\textrm{ms}} \sim \int dt \left(\frac{\dot{V}^2(t)}{V(t)} - 4 \Lambda V(t)\right),
\end{equation}
where $\Lambda$ is the cosmological constant from the Euclidean Einstein-Hilbert action. The first term containing the time derivative $\dot{V}(t)$ is a kinetic contribution, and the term containing the cosmological constant forms the potential. A straightforward discretization of this continuum effective action is 
\begin{equation}
\label{eq:minisuper-disc}
    \seff = \sum_{t} \frac{1}{\Gamma} \frac{(n_{t+1} - n_t)^2}{n_{t+1} + n_t} + \lambda n_t ,
\end{equation}
where $\Gamma$ and $\lambda$ are constants we can fit to the numerical results. We denote the discrete kinetic term (proportional to $1/\Gamma$) by $K[n_t, n_{t+1}]$ and the discrete potential term (proportional to $\lambda$) by $U[n_t]$.

Computing the matrix $P_{tt'}$ using the ansatz \eqref{eq:minisuper-disc} for $\seff$, we see that it is nonzero only on the tridiagonal. The subdiagonal and superdiagonal are equal since the matrix is symmetric by definition. Furthermore, the sub- and superdiagonals only receive a contribution from the kinetic term $K[n_t, n_{t+1}]$. The diagonal receives contributions from both the kinetic term and the potential. Let us define the \emph{kinetic coefficient} $k_t$ and the \emph{potential coefficient} $u_t$ as follows:
\begin{align}
    k_t \equiv -P_{t(t+1)} &= \frac{1}{\Gamma} \frac{2}{\langle n_t \rangle + \langle n_{t+1} \rangle} \\
    u_t &=P_{tt} - k_t - k_{t-1} .
\end{align}
By measuring the covariance matrix in our computer simulations and inverting it, we can therefore extract kinetic coefficients $k_t$ from the off-diagonals and subsequently compute the potential coefficients $u_t$ by subtracting the potential coefficients from the diagonal. Since the systems we simulate are translationally invariant\footnote{This includes the case of spatial genus 0 in this setup, as explained earlier.} in the time direction, the values within each (sub-)diagonal are approximately equal and it makes sense to define the averaged coefficients
\begin{align}
    k &= -\frac{1}{2T} \sum_t (\langle P_{t(t+1)}\rangle + \langle P_{t(t-1)}\rangle), \\
    u &= -2k + \sum_t \langle P_{tt}\rangle.
\end{align}
The matrix indices in the summation for $k$ are taken modulo $T$ to account for the periodicity in the time direction. The measured $P_{tt'}$ (after subtraction of the volume-fixing contribution $2\epsilon$) then approximately takes the form of the following Toeplitz matrix:
\begin{equation}
P_{tt'} = C^{-1}_{tt'} =
\begin{pmatrix}
2k+u      & -k    & 0 & 0 & \dots & -k \\
-k        & 2k+u  & -k        & 0 & \dots & 0 \\
0 & -k    & 2k+u      & -k        & \dots & 0 \\
\dots & \vdots    & \vdots & \vdots &\ddots  & 0 \\
-k     & 0  & 0 & 0 & \dots & 2k+u
\end{pmatrix}
.
\end{equation}

Our simulations for a range of distinct system volumes then provide us with a series of kinetic and potential coefficients $k(\bn), u(\bn)$ --- where $\bn = \aavn = \tn / T$ --- from which we can reconstruct the effective action. We performed measurements for the target volumes $\tn \in \{2,3,4,5,6,7,9,10,20,40,60,80\} \times 1000$, for the three ensemble types II, IIA, and III. The values of the spatial genus where we performed simulations for all three ensemble types are $g \in \{0,1,5,15\}$, and we carried out an extra set of simulations exclusively for the type III ensemble with $g \in \{0,1,2,3,4,5,6,7,8,9,14,15,20,30\}$. All simulations were performed at $k_0 = 5.0$. We collected on the order of half a million volume profiles for each set of input parameters, with $10 \cdot\tn$ attempted geometric moves (which we collectively refer to as one \emph{sweep}) in between subsequent volume profiles. For the smaller system volumes we obtained several million volume profiles since the sweeps are significantly less time-consuming.

In what follows, we first present our results for the kinetic term, which allow for a straightforward extraction of the kinetic constant $\Gamma$. It turns out that our results for the potential term are more difficult to interpret, so we postpone this discussion to the last part of this section.

\subsubsection{The kinetic term}
In order to fix $K[n_t, n_{t+1}]$ we need to determine the kinetic constant $\Gamma$ as it appears in \eqref{eq:minisuper-disc}. This can be accomplished by performing a linear fit for the parameters $a, \Gamma$ to our results for the kinetic coefficients $k(\bn) $, using the following relation:
\begin{equation}
    \frac{1}{k(\bn)} = a + \Gamma \, \bn
    \label{eq:kin-constant}
\end{equation}
where we introduced the additive shift parameter $a$ to account for the fact that we do not expect small system sizes to provide a proper picture of the continuum behavior of the system.

\begin{figure}[ht!]
\centering
\includegraphics[width = 0.8\textwidth]{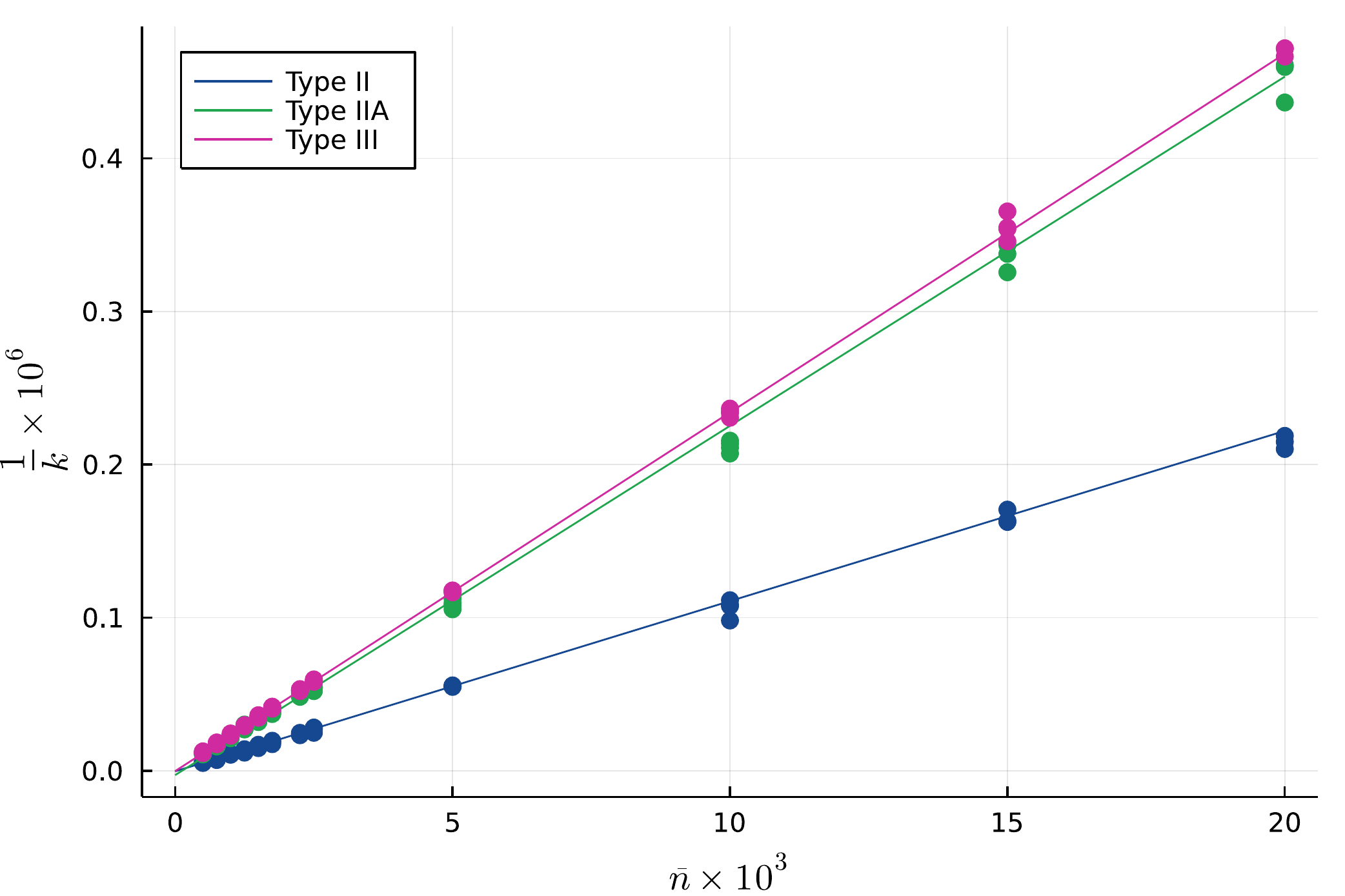}
\caption{The inverse kinetic coefficients $1/k$ as a function of the volume per slice $\bn$. The data points are color grouped by ensemble types II (blue), IIA (green), and III (purple). The spatial genera used in the simulations are $g \in \{0, 1, 5, 10,15\}$. The straight lines are linear fits to the genus 1 data obtained for the three distinct ensemble types. (Error bars are on the order of the dot size.)}
\label{fig:kinetic-coefficient}
\end{figure}

We present our measurement results for the genus- and ensemble-type-dependent inverse kinetic coefficients $1/k$ as a function of the average volume per slice $\bn = \tn / T$ in Fig. \ref{fig:kinetic-coefficient}. The shown data points are for the genera $g \in \{0,1,5,15\}$ and all three types of ensembles. We indeed observe a linear dependence of $1/k$ on $\bn$, justifying the fit proposed in \eqref{eq:kin-constant}. It is interesting to note that the slope parameter of this relation seems to be largely independent of the spatial genus. On the other hand, the choice of ensemble type has a clear effect on the slope, with especially the type II ensemble standing apart from the other two. This leads us to define one kinetic constant $\Gamma_s$ for every distinct ensemble type $s$, where we fix $\Gamma_s$ to be the best fit slope parameter in \eqref{eq:kin-constant} for the data obtained for the system with spatial genus 1. The resulting values for $\Gamma_s$ can be found in Table \ref{tab:param-fits}.
\begin{table}[ht!]
\centering
\begin{tabular}{ |c|c| }
\hline
 Ensemble type & $\Gamma (\times 10^3)$ \\ 
 \hline
 II & $11.1(2)$ \\  
 IIA & $22.8(5)$ \\
 III & $23.4(2)$ \\
 \hline
\end{tabular}
\caption{Best fit values for the kinetic constant $\Gamma$ for the three ensemble types used in our simulations.}
\label{tab:param-fits}
\end{table}

Interestingly, the resulting constants for the type IIA and III ensembles slightly fall within each others margins of error, raising the question whether they may in fact be equal. Judging from Fig. \ref{fig:triangulation-types}, it is perhaps surprising that this approximate equality occurs between the type IIA and III ensembles, while the structure of the spatial slices of the type IIA ensemble resembles that of the type II ensemble more. However, we should point out that all measurements were performed at $k_0 = 5.0$, regardless of the ensemble type. Recalling Fig. \ref{fig:op2}, we see that the location of the critical point $\kgc$ is different for all three ensembles, so that the value $k_0 = 5.0$ we singled out is at a different distance from the transition for simulations performed with distinct ensembles. It was conjectured in \cite{ambjorn2001nonperturbative} that the value of $k_0$ inside the de Sitter phase merely sets the length scale of the system, which should be taken into account when interpreting absolute values of the kinetic constant $\Gamma$ that we determined using our measurements. As long as we lack a better understanding of the effect of adjusting $k_0$ within the de Sitter phase, direct comparisons of $\Gamma$ between distinct ensembles are likely not meaningful.

\subsubsection{The potential}
We now turn our attention to the potential coefficients $u$, and how these can be represented in the potential term $U[n_t]$ of the effective action. In fact, there is an important subtlety related to our ansatz \eqref{eq:minisuper-disc}. We assumed that
\begin{equation}
    U[n_t] = \lambda n_t.
\end{equation}
However, as a consequence of the relation
\begin{equation}
    u(\bn) = \left. \frac{\partial^2 U[n_t]}{\partial n_t^2}\right|_{n_t=\bn},
\end{equation}
we may expect all potential coefficents $u$ to vanish. As we will see below, this is not the case in our numerical results. However, we can still find a good match if we consider the following slightly adapted ansatz $\tilde{U}[n_t]$ for the potential:
\begin{equation}
    \tilde{U}[n_t] = \lambda n_t^{1+\epsilon},
\end{equation}
where $\epsilon$ is a small, positive real number. We then find that
\begin{equation}
    u(\bn) = \lambda \epsilon (1+\epsilon) \bn^{-1+\epsilon},
\end{equation}
which proves to be a significantly better match with our data.

We measured the covariance matrices for all three ensemble types II, IIA, and III for spatial genus $g \in \{0,1,5,15\}$, and used these to determine the potential coefficients $u$. The potential coefficients we measured are small compared to the kinetic coefficients $k$. Since $u$ is obtained by subtracting $2k$ from the diagonal of $P_{tt'}$, the error bars for these potential coefficients are relatively large --- generally on the order of 10\% of the average value for the type III ensemble.\footnote{The error bars were determined using a bootstrap resampling procedure.} The relative errors were even larger for the type II and IIA ensembles, and we therefore chose to restrict our attention to the type III ensemble for a more detailed study of the potential. It is, however, interesting to note that the measured potential coefficients for a fixed value of the spatial genus do not seem to depend on the type of ensemble, in the sense that the error bars of corresponding data types roughly overlap. However, since the errors for the type II and IIA ensembles were large, the best we can do at this stage is to view this as a conjecture.

\begin{figure}[ht!]
\centering
\includegraphics[width = 0.8\textwidth]{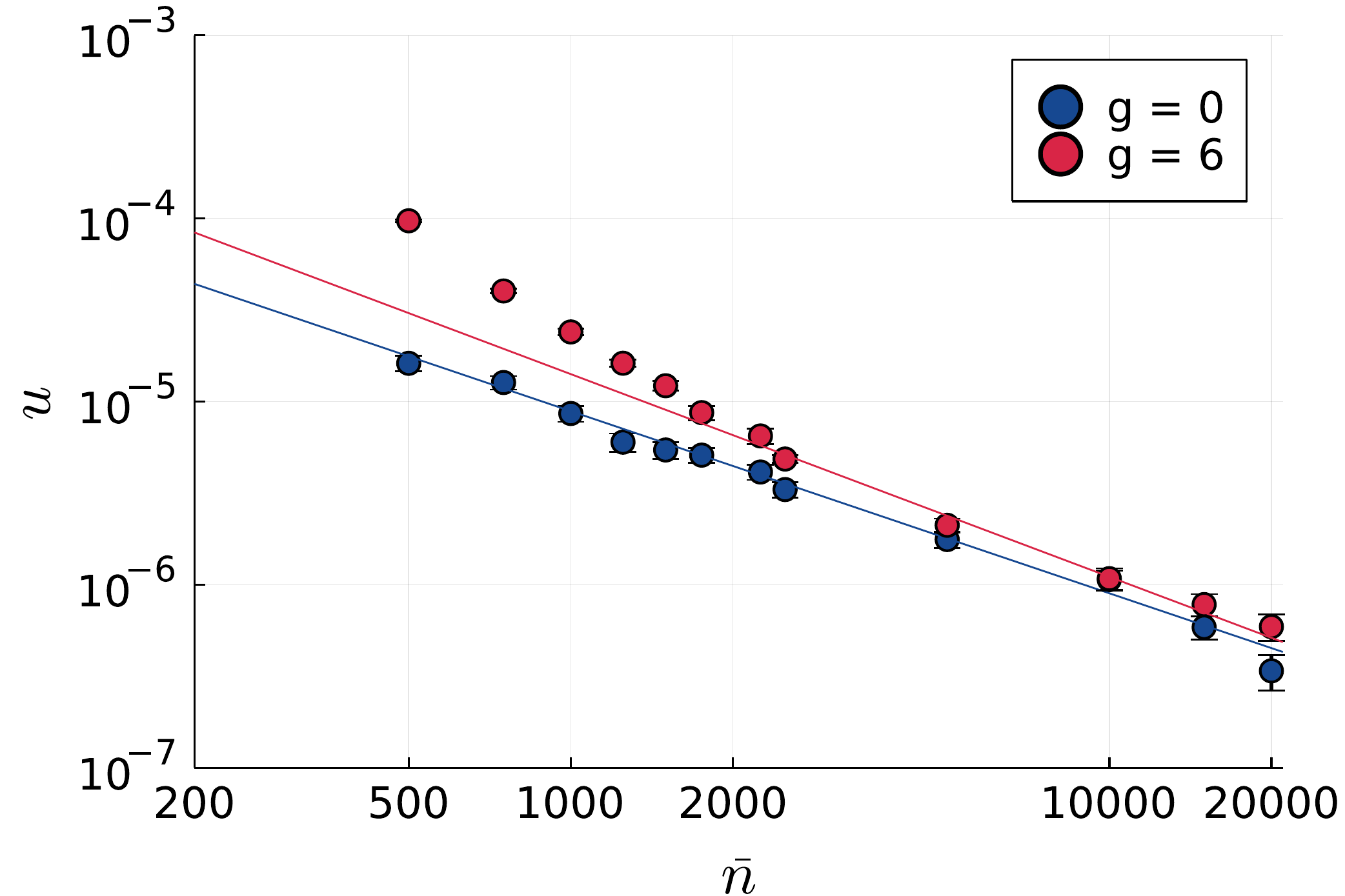}
\caption{A log-log plot of the potential coefficients $u$ as a function of the volume per slice $\bn$ for the type III ensemble with spatial genus $g=0,6$. The straight lines are best fits of the form $u = c \cdot \bn^\beta$, where only the data points with $\bn \geq 1750$ are taken into account for the fit.}
\label{fig:pot-coeff-0-6}
\end{figure}

For our more in-depth analysis of the type III ensemble, we measured the covariance matrices for spatial genus $g \in \{0,1,2,3,4,5,6,7,8,9,14,15,20,30\}$ and subsequently extracted the potential coefficients $u(\bn)$. As an example of the general pattern we observe in our measurements, we present the results for spatial genus $g=0$ and $g=6$ in Fig. \ref{fig:pot-coeff-0-6}. The data points are shown on a doubly logarithmic scale, so that a power law relation $u \sim \bn^\beta$ would appear as a straight line. We observe that the results for $g=0$ approximately lie along a straight line for the full range of system volumes $\tn = T \cdot \bn$ used in the simulations. The results for $g=6$ deviate from this straight line for small system volumes, but the data points for $\bn \geq 2250$ roughly follow the same pattern as the results for $g=0$. For convenience, we plotted best fits of the form $c \cdot \bn^\beta$, where only the data points for $\bn \geq 2250$ were taken into account for the fits. The corresponding best fit values for the exponent are $\beta=-0.99(18)$ for $g=0$ and $\beta=-1.07(15)$ for $g=6$. Repeating this procedure for all $g \leq 7$ leads to a similar result, where the best fit exponent $\beta$ is approximately equal to $-1$ within numerical error when fitting to the data points for which $\bn \geq 2250$. The larger the genus, the stronger the deviation from the straight line is for the smaller volumes. For the values $g > 7$ that we investigated, this pattern disappears, and the best fit exponent for the large-$\bn$ data points swiftly decreases with genus.

Our interpretation of these findings is the following. The best fit values for the exponent $\beta$ in the power law relation $u \sim \bn^\beta$ are close to $-1$ for low values of the spatial genus, which suggests behavior consistent with the ansatz $\tilde{U}[n_t] = \lambda n_t^{1+\epsilon}$, with $\epsilon$ a positive real number close to zero. We understand this as evidence supporting the conjectured potential term $U[n_t] = \lambda n_t$ obtained from a minisuperspace treatment of the quantum gravitational path integral. The fact that we observe increasing deviations from this behavior at small volumes as the spatial genus increases may be attributable to discretization artefacts. This is supported by our determination of the effective minimal slice sizes in Appendix \ref{app:slice-vol-dists}, where we found that this minimal size increases monotonically with the spatial genus. Deviations from continuum behavior can therefore be fully expected when typical slice sizes are too close to their minimum. It is likely that this is also the reason that we did not observe $u \sim \bn^{-1+\epsilon}$ behavior for large values of the genus, since discretization artefacts may play a significant role even at the larger volumes $\bn \geq 2500$ where we perform the fit.

\begin{figure}[ht!]
\centering
\includegraphics[width = 0.48\textwidth]{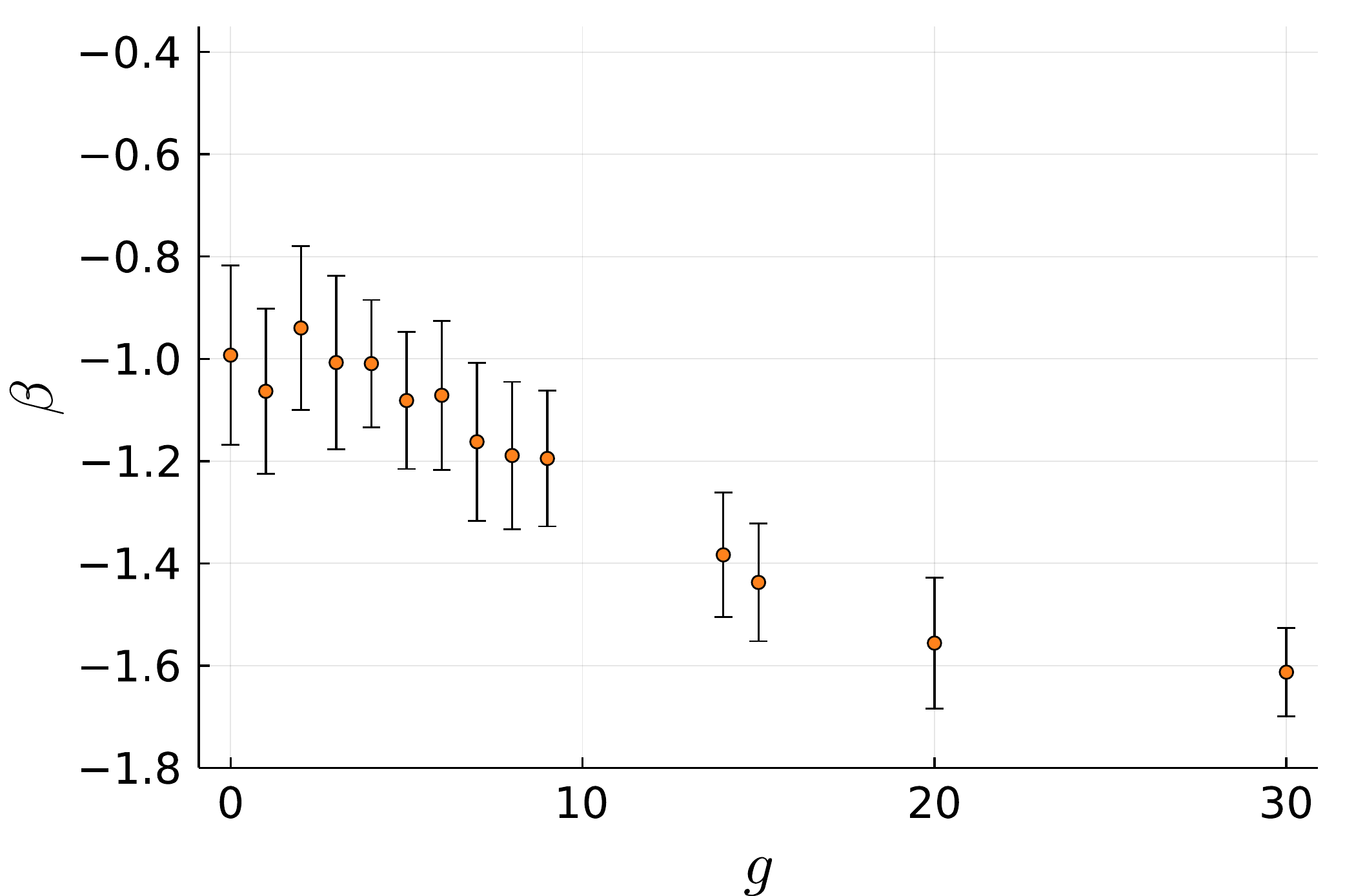} 
\hfill
\includegraphics[width=0.48\textwidth]{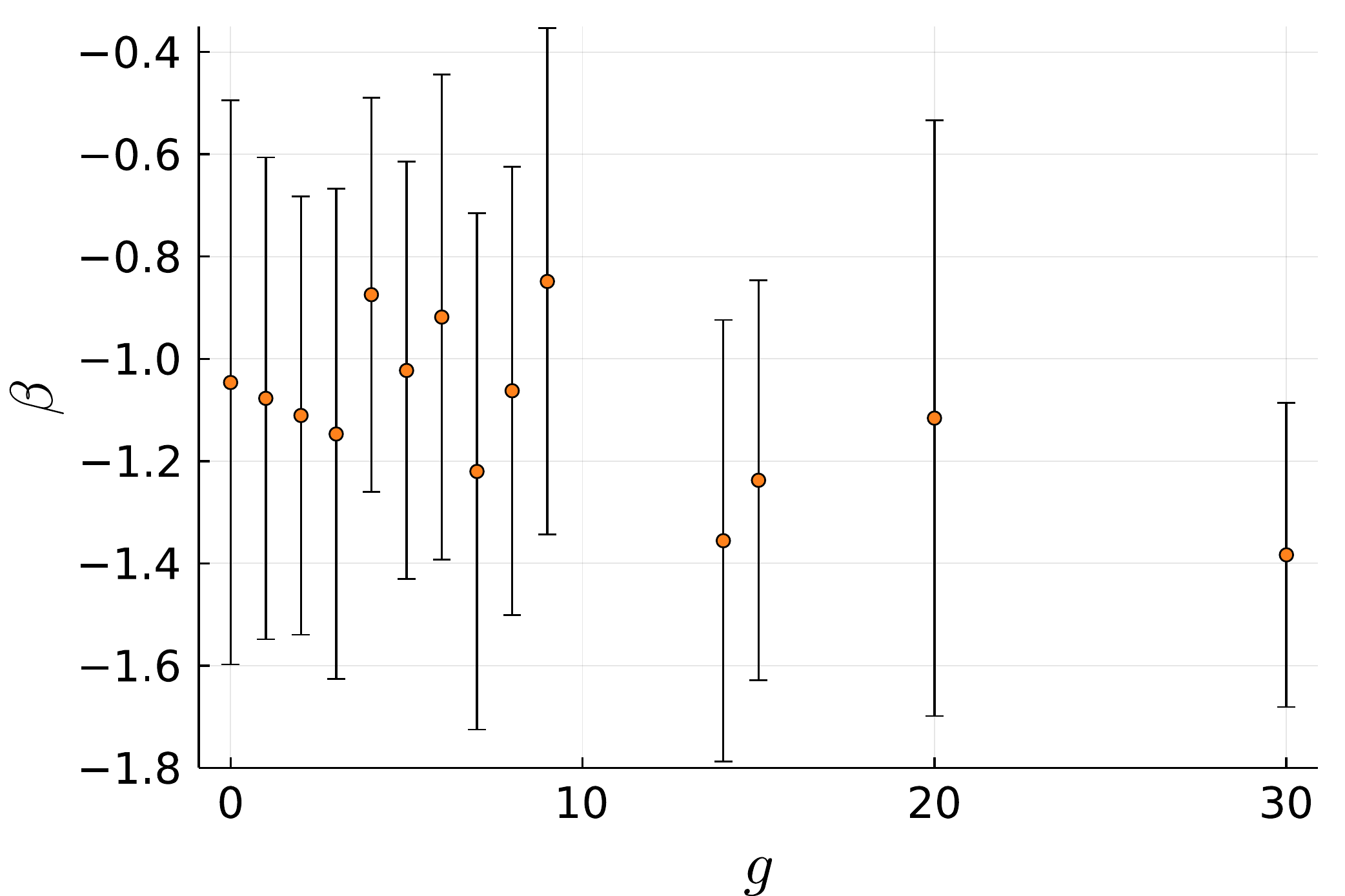}
\caption{Best fit parameters with associated 95\% confidence intervals for the exponent $\beta$ in the ansatz $u \sim \bn^\beta$, obtained using two distinct fit ranges. The slice volumes used for the fits in the leftmost plot are $\bn \in \{2250, 2500, 5000, 10000, 15000, 20000\}$, while for the rightmost plot we used $\bn \in \{5000, 10000, 15000, 20000\}$.}
\label{fig:pot-fits}
\end{figure}

In order to demonstrate this behavior, we plot the best fit parameters for two distinct fit ranges in Fig. \ref{fig:pot-fits}. For the leftmost plot, we performed the fit for the six data points in the range $\bn \in [2250, 20000]$, while for the rightmost plot we used the four data points in the range $\bn \in [5000, 20000]$. The error bars are 95\% confidence intervals. We see that the best fit exponents obtained in the wider range $\bn \in [2250, 20000]$ are approximately consistent with the value $\beta = -1$ for $g \leq 7$, but that this value quickly drops off for larger values of $g$. All error margins are of the order $\sim 0.2$. When we fit to the more narrow range $\bn \in [5000, 20000]$, we see that the error margins increase to a size of the order $\sim 0.5$, making it difficult to draw strong conclusions from these numbers. However, for this fit range almost all best fit exponents are consistent with the value $\beta = -1$, which hints at a potential term of the form $U[n_t] = \lambda n_t$ suggested by the minisuperspace action. At this stage, the best we can do is to consider this as an encouragement to perform more detailed measurements with increased resolution in the regime of large system volumes. This is, however, beyond the scope of the current work.

\section{Discussion}
\label{sec:discussion}
We analyzed aspects of (2+1)-dimensional CDT using Monte Carlo simulations, on the one hand studying the behavior of the model as the simplicial manifold conditions are relaxed, and on the other hand investigating the effect of changing the topology of the spatial slices. We find that both these variations do not affect the qualitative structure of the phase space, although we do observe certain differences on a quantitative level: the location of the phase transition shifts under these changes, possibly implying a change of scale in the coupling constants. Furthermore, the volume scale at which discretization artefacts manifest themselves is also affected by both the ensemble type and spatial genus. However, our data indicate that the phase transition remains first-order regardless of ensemble type or spatial genus, and that certain global geometric properties of the two phases on both sides of this transition are also left untouched. That is, the volumes of neighbouring spatial slices in the degenerate phase ($k_0 > \kgc$) are largely uncorrelated, whereas such neighbouring slices in the de Sitter phase ($k_0 < \kgc$) tend to have similar volumes. To reiterate, this behavior is independent of the three different geometric ensembles we investigated on the one hand, and of the topological genus of the spatial slices on the other hand.

These observations can be interpreted as evidence that mildly relaxing the simplicial manifold conditions does not affect the universal behavior of the system in a continuum limit, similar to what has been analytically proven in the setting of two-dimensional Euclidean Dynamical Triangulations. We should point out that we have only partially relaxed the simplicial manifold conditions on the two-dimensional spatial \emph{slices} of the (2+1)-dimensional Lorentzian model, and that we did not attain maximally degenerate \emph{spacetime} triangulations. Extending the (2+1)-dimensional CDT model to maximally degenerate configurations involves significantly larger ensembles than the ones we investigated, and our results do \emph{not} imply that this fully extended model exhibits the same universal behavior. In fact, an analysis of more degenerate ensembles using matrix models has indicated that the critical point $\kgc$ shifts to infinity when one considers fully degenerate triangulations \cite{ambjorn2001lorentzian}, so that only the de Sitter phase survives. It would be interesting to further investigate this issue by extending our simulation code to be compatible with such degenerate geometries.

Furthermore, we studied whether the volumes of the spatial slices (and their fluctuations) in the de Sitter phase of the model can be described by a semiclassical effective action. We found a strong signal of a quasi-local kinetic term that describes the coupling between volumes of neighboring slices, in all three ensemble types and for all values of the spatial genus that we investigated. Extracting a potential term proved to be more challenging, partly since the potential coefficients measured in our simulations were subject to relatively large numerical noise --- especially so for the type II and IIA ensembles. We therefore restricted our attention to the type III ensemble, where we concluded that our results for spatial genus $g \lessapprox 7$ are at least \emph{consistent} with a linear potential term $U[n_t] = \lambda n_t$ in the effective action, which is also the functional form of the potential suggested by the minisuperspace action for 3D gravity. Our results for $g \gtrapprox 7$ are more difficult to interpret, but our measurements of the distributions of spatial volumes at distinct values of $g$ in Appendix \ref{app:slice-vol-dists} make it plausible that discretization artefacts are too dominant at our typical system sizes to put much trust in the results at high $g$.

When interpreting the results for the kinetic constant $\Gamma$, we pointed out that direct comparisons of these constants between distinct ensembles are likely not meaningful. The reason is that the location of the critical point $\kgc$ depends on the ensemble, and the value $k_0 = 5.0$ at which we determined the covariance matrices therefore has a different distance from the transition in each of the ensembles. Furthermore, the scale of the associated pseudocritical lines $\kcc(k_0)$ and order parameter $\optwo(k_0)$ is also different between the phases. Building upon the conjecture from \cite{ambjorn2001nonperturbative} that the value of $k_0$ within the de Sitter phase merely sets the overall physical scale of the system, it would be interesting to measure covariance matrices at other values of $k_0$ inside this phase, and to investigate how this affects the resulting values of the kinetic constant $\Gamma$. Such results could perhaps be used to pin down the functional dependence of this emergent length scale on the bare coupling constant $k_0$.

A natural extension to this work would be to make a more detailed study of the potential term in the effective action. This can, for example, be achieved by measuring the covariance matrices for a range of system volumes such that $\bn \in [5000, 10000]$ with increased resolution, so that we can extract more precise estimates of the exponent $\beta$ in the relation $u \sim \bn^\beta$ for values of the spatial genus $g \lessapprox 10$. This would serve to provide stronger evidence that the leading potential term in the effective action is indeed linear in $\bn$, which makes it likely that changing the spatial genus does not affect the leading behavior of the potential --- that is, for the range of $g$ under consideration.

\vspace{0.5cm}
\noindent{\bf Acknowledgments.}
This work was partly supported by a Projectruimte grant of the Foundation for Fundamental Research on Matter (FOM, now defunct), financially supported by the Netherlands Organisation for Scientific Research (NWO). We would like to thank Jan Ambj\o rn, Timothy Budd, and Renate Loll for useful discussions and comments on the manuscript.

\begin{appendices}
\section{Slice volume distributions}
\label{app:slice-vol-dists}
In this Appendix, we present distributions of the volumes of the spatial slices in the degenerate phase of 3D CDT at distinct spatial genus, determined with the help of computer simulations. The motivation for this is to get an estimate of the effective minimal sizes of the spatial triangulations as they appear in the simulations performed for the main body of this work. Having these estimates allows us to form an impression of the relative size of the systems used in our main simulations, in comparison to the configurations of minimal size. This, therefore, gives a rough handle on the scale at which we can expect lattice discretization artefacts to play a major role, which we can subsequently avoid when collecting data for our main results. 

\begin{figure}
    \centering
    \includegraphics[width=0.45\textwidth]{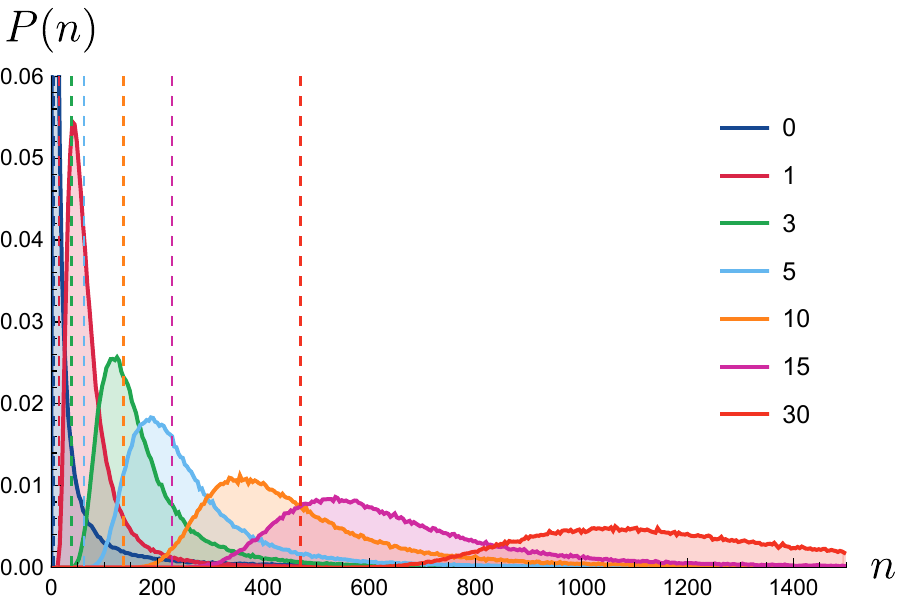}
    \hfill
    \includegraphics[width=0.45\textwidth]{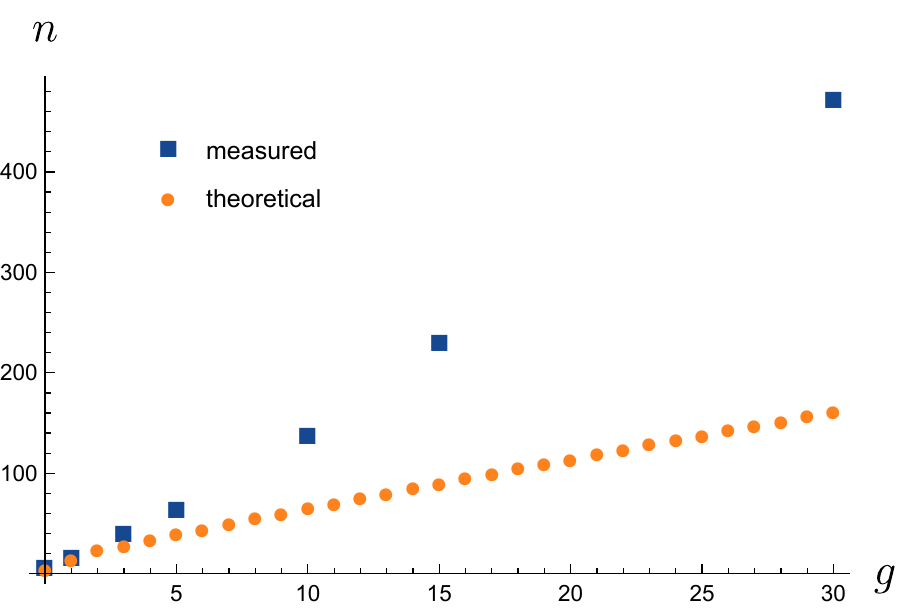}
    \caption{Histograms of slice volumes (left) and minimal slice sizes (right) in the type III CDT ensemble with $T=10$ time slices and system volume $\tilde{N}_{31}=20k$, for distinct values of the spatial genus $g \in \{0,1,3,5,10,15,30\}$. The vertical dashed lines in the histogram plot indicate the minimal slice sizes for each value of $g$. The dashed line in the rightmost plot shows the theoretical lower bound $\delta(\mathbb{T}^g)$ for the number of triangles required to triangulate an oriented genus-$g$ surface. Note: the distribution for $g=0$ peaks near $0.26$ and is partially cut off from the plot.}
    \label{fig:volume-distributions}
\end{figure}

In the context of nondegenerate triangulations (corresponding to spatial universes in our type III ensemble) it is known \cite{jungerman1980minimal} that the minimal number of triangles $\delta(\mathbb{T}^g)$ required to triangulate an orientable surface $\mathbb{T}^g$ of genus $g \leq 3$ equals
\begin{equation}
    2 \left\lceil \frac{7+\sqrt{1+48 g}}{2} \right\rceil + 4(g-1),
\end{equation}
where $\lceil x \rceil$ is the smallest integer $\leq x$. Furthermore, we have that $\delta(\mathbb{T}^0)= 4, \delta(\mathbb{T}^1) = 14, \delta(\mathbb{T}^2) = 24$, where we understand $\mathbb{T}^0$ to be the 2-sphere $S^2$. For the largest value of the spatial genus that we investigated, $g=30$, we find that $\delta(\mathbb{T}^{30}) = 162$. When determining the best fit parameters in extracting the kinetic and potential terms of the effective action, we only took into account the data points where the slice volumes exceed 1250 triangles, which is well away from this minimal size. However, we should take into account that $\delta(\mathbb{T}^g)$ is a lower bound, and it is not guaranteed that this bound is saturated in our computer simulations. It may be the case that the embedding of the two-dimensional triangulations in the three-dimensional simplicial manifold results in further constraints that push up the minimum required number of triangles of a spatial slice of genus $g$, or it may simply be highly unlikely for the random walk through the space of triangulations to attain this minimum.\footnote{For example, it is possible that the only path leading to the minimal configuration using our prescribed Monte Carlo updates passes through a highly symmetrical spatial triangulation that is much larger than the minimal one, and that a precise sequence of moves subsequently needs to be followed in order to reach this minimal configuration.} Therefore, it is appropriate to take a more realistic approach and investigate the actual sizes of minimal triangulations encountered in our simulations. To this end, we measured volume profiles in the \emph{degenerate} phase of 3D CDT. The reason for studying the degenerate phase here is that the spatial volumes in this phase are not distributed equally over the slices, but rather tend to peak in a small number of slices. This can be seen for spatial genus 0 and 1 in Fig. \ref{fig:sphere-torus-vp}. As a result, most slices have small volumes, making it more likely that they approach the minimum possible size. We ran simulations for several values of the genus $g \in \{0,1,3,5,10,15,30\}$ at target system volume $\tilde{N}_{31} = 20k$ and $T=10$ time slices, collecting on the order of $10^3$ volume profiles for each value of $g$. We show the resulting histograms of spatial volumes (left) and the size of the smallest encountered slices (right) for each genus in Fig. \ref{fig:volume-distributions}. We see that the volume distribution for the case of a spatial sphere $(g=0)$ is strongly peaked at the minimal slice size $n_t = 4$, and that the distributions gradually spread out as the genus increases. Furthermore, it is clear that the size of the minimal slices increases approximately linearly in $g$. Although these values are strictly larger than the theoretical lower bound $\delta(\mathbb{T}^g)$ for the number of triangles in a minimal triangulation of a genus-$g$ surface, they differ by a multiplicative factor on the order of $\sim 2.5$. 
We point out that the minimal slice sizes found using this method do not provide us with a lower bound above which our results are guaranteed to be useful. We should rather understand these as typical sizes for which discretization artefacts are almost guaranteed to influence the results, which serves as a warning sign that we cannot put too much trust in simulations where slice sizes are likely to approach these numbers. It is therefore advisable to work with systems where typical slice sizes are much larger than this minimum if we want to probe continuum behavior of the model.

\end{appendices}

\newpage

\bibliographystyle{JHEP}
\bibliography{3d-eff-act}

\end{document}